**Effectiveness and Safety of Selective IL-23 Receptor Antagonists in Moderate to Severe Ulcerative Colitis: A Systematic Review, Meta-Analysis and Trial Sequential Analysis**

**Eficácia e Segurança dos Antagonistas Seletivos do Receptor de IL-23 na Retocolite Ulcerativa Moderada a Grave: Uma Revisão Sistemática, Meta-Análise e Análise Sequencial de Ensaios**

Wellgner Fernandes Oliveira Amador[1], Isabelle Castro Vitor[2], Milena Ramos Tomé, MD[3], Diogo Delgado Dotta, MD[4], Rodrigo V Motta, MD[5]

[1]Department of Medicine, Federal University of Campina Grande, Cajazeiras, Paraíba, Brazil. E-mail: wellgnerfernandes@gmail.com. ORCID: 0009-0002-3186-2589.

[2]Department of Medicine, Federal University of the State of Rio de Janeiro, Rio de Janeiro, Brazil. Email: isabellecastro94@gmail.com. ORCID: 0009-0009-8698-8737.

[3]Department of Medicine, Federal University of Campina Grande, Cajazeiras, Paraíba, Brazil. Email: milenaramos19@gmail.com. ORCID: 0009-0007-9381-0684.

[4]Department of Gastroenterology and Hepatology, Hospital das Clínicas of Medical School, University of São Paulo. Email: diogo.delgadodotta@gmail.com. ORCID: 0000-0002-3790-5079.

[5]Translational Gastroenterology and Liver Unit, Nuffield Department of Medicine, Experimental Medicine Division, University of Oxford. Email: rodrigo.vieiramotta@ndm.ox.ac.uk. ORCID: 0000-0002-6633-9133.

**Short title**: IL-23 antagonists in Ulcerative Colitis

**Word count**: 3,819 words.

**Figures**: 3.

**Tables**: 2.

**Corresponding author**: Wellgner F. O. Amador. E-mail: wellgnerfernandes@gmail.com. Address: Dep. José Edmur Estrela Street, number 166, São Joao do Rio do Peixe, Paraiba, Brazil. ORCID: 0009-0002-3186-2589.

**Statements and Declarations**

**Ethics approval**

Not applicable.

**Conflict of interest**

The authors declare they have no financial interests.

**Funding Sources**

This research did not receive any specific grant from funding agencies in the public, commercial, or not-for-profit sectors.

**Contributor Roles**

Wellgner Fernandes Oliveira Amador - Conceptualization, Data curation, Formal analysis, Investigation, Methodology, Resources, Project administration, Visualization, Writing – original draft, Writing – review & editing

Isabelle Castro Vitor - Investigation, Methodology

Milena Ramos Tomé, MD - Investigation

Diogo Delgado Dotta, MD - Visualization, Writing – review & editing

Rodrigo V Motta, MD - Conceptualization, Investigation, Methodology, Supervision, Validation, Visualization, Writing – original draft, Writing – review & editing

**Abstract**

***Background and Aim***: Selective interleukin-23 receptor antagonists (IL-23RA) show promise for treating moderate to severe ulcerative colitis (UC), but their efficacy and safety are not fully understood. Hence, this systematic review and meta-analysis assess the effectiveness and safety of IL-23RA in UC.

**Methods**: A systematic search of PubMed, Embase, Cochrane, and ClinicalTrials.gov was performed in December 2024. Randomized controlled trials (RCTs) comparing IL-23RA to placebo in moderate to severe UC were included. Outcomes included clinical and endoscopic remission, response rates, and adverse events (AEs). Risk ratios (RR) and mean differences (MD) with 95% confidence intervals (CI) were pooled using a random-effects model in R software.

***Results***: Nine RCTs (3,808 patients in the induction phase; 1,734 in the maintenance phase) were analyzed. IL-23RA enhanced clinical remission (induction: RR 2.63; 95% CI 2.05–3.36; maintenance: RR 1.99; 95% CI 1.63–2.44; all $p < 0.01$) and endoscopic remission (induction: RR 2.36; 95% CI 1.70–2.20; maintenance: RR 1.96; 95% CI 1.63–2.37; all $p < 0.01$). IL-23RA reduced serious AE in the induction phase (RR 0.40; 95% CI 0.27-0.69; $p < 0.01$), while there was no difference during maintenance (RR 0.75; 95% CI 0.31-1.84; $p = 0.53$). No differences were observed in overall AEs or specific AEs like headache or nasopharyngitis. Trial sequential analysis confirmed sufficient sample size for clinical endpoints.

***Conclusions***: IL-23RA showed superior effectiveness and similar safety when compared to placebo in UC.

***Keywords***: Ulcerative colitis, Interleukin-23, Effectiveness, Safety.

**Resumo**

**Contexto e Objetivo:** Antagonistas seletivos do receptor da interleucina-23 (IL-23RA) demonstram potencial no tratamento da retocolite ulcerativa (RCU) moderada a grave, mas sua eficácia e segurança ainda não estão completamente esclarecidas. Assim, esta revisão sistemática e meta-análise avaliou a efetividade e segurança dos IL-23RA na RCU.

**Métodos:** Foi realizada uma busca sistemática nas bases PubMed, Embase, Cochrane e ClinicalTrials.gov em dezembro de 2024. Foram incluídos ensaios clínicos randomizados (ECRs) que compararam IL-23RA com placebo em pacientes com RCU moderada a grave. Os desfechos incluíram remissão clínica e endoscópica, taxas de resposta e eventos adversos (EAs). Razões de risco (RR) e diferenças médias (DM) com intervalos de confiança (IC) de 95% foram combinadas utilizando um modelo de efeitos aleatórios no software R.

**Resultados:** Foram analisados nove ECRs (3.808 pacientes na fase de indução; 1.734 na fase de manutenção). O uso de IL-23RA aumentou a remissão clínica (indução: RR 2,63; IC 95% 2,05–3,36; manutenção: RR 1,99; IC 95% 1,63–2,44; todos p < 0,01) e a remissão endoscópica (indução: RR 2,36; IC 95% 1,70–2,20; manutenção: RR 1,96; IC 95% 1,63–2,37; todos p < 0,01). Os IL-23RA reduziram os EAs graves na fase de indução (RR 0,40; IC 95% 0,27–0,69; p < 0,01), sem diferença significativa na fase de manutenção (RR 0,75; IC 95% 0,31–1,84; p = 0,53). Não foram observadas diferenças nos EAs gerais ou em EAs específicos como cefaleia ou nasofaringite. A análise sequencial de ensaios confirmou amostra suficiente para os desfechos clínicos.

**Conclusões:** Os IL-23RA demonstraram eficácia superior e segurança semelhante ao placebo em pacientes com RCU.

**Palavras-chave:** Retocolite ulcerativa, Interleucina-23, Eficácia, Segurança.

## 1. Introduction

Ulcerative colitis (UC) is a chronic inflammatory condition of the colon characterized by periods of exacerbation and remission, significantly impacting patients' quality of life (1–4). In North America, the prevalence of UC is 0.4% and it affects approximately 1.5 million people (4). While tumor necrosis factor (TNF) inhibitors, including infliximab and adalimumab, are widely used as first-line therapies, nearly one-third of patients fail to achieve clinical remission (5). Additionally, other available treatments, such as immunomodulators and corticosteroids, are often constrained by limited effectiveness and a risk of adverse events, underscoring the need for more effective and safer therapeutic options (1,6).

While the pathogenesis of UC remains complex, it is increasingly understood that interleukin-23 (IL-23) plays a critical role in driving the pro-inflammatory responses central to disease progression (7–9). As a result, selective IL-23 receptor antagonists (IL-23RA) have emerged as a promising class of therapies, targeting a key cytokine pathway involved in inflammation while minimizing off-target effects (7,9). These agents have shown effectiveness in inducing and maintaining remission in moderate to severe UC, with a favorable safety profile (10,11).

Despite promising results from randomized controlled trials (RCTs), key gaps persist in understanding the comparative effectiveness, long-term safety, and class-wide consistency of selective IL-23RA in moderate to severe UC. Previous network meta-analyses (NMA) have evaluated biologics and small molecules for UC, but these primarily emphasized efficacy outcomes and did not focus specifically on the IL-23 inhibitor class (12,13). To address these gaps, this systematic review and meta-analysis evaluates both the effectiveness and safety of IL-23RA, incorporating trial sequential analysis (TSA) to strengthen result reliability, bridge gaps in the literature, and provide actionable evidence to optimize treatment strategies for UC patients.

## 2. Methods

This systematic review followed the recommendations of the Cochrane Collaboration (14) and the Preferred Reporting Items for Systematic Reviews and Meta-Analysis (PRISMA) guidelines (15), including the design, implementation of the steps, analysis, and description of the results. The study protocol was registered in the International Prospective Register of Systematic Reviews (PROSPERO) under registration number CRD42024618348.

### 2.1 Search strategy

A systematic search on PubMed (MEDLINE), Embase, Cochrane Central, and Clinical Trial databases was conducted on December 25, 2024. The following medical subject heading terms have been included: 'ulcerative colitis', 'il-23', 'interleukin-23', 'guselkumab', 'tildrakizumab', 'mirikizumab', 'ustekinumab', 'risankizumab', 'randomized controlled trial', 'controlled clinical trial', 'randomized', 'placebo', 'drug therapy', 'randomly', 'trial', and 'groups'. The search strategy is detailed in **Supplementary Table S1**.

### 2.2 Data extraction

After removing duplicates, two authors (I.V. and W.A.) screened the titles and abstracts, independently evaluating the full-text articles for inclusion based on pre-specified criteria. Discrepancies were discussed and resolved by a third reviewer (M.T.). Data extraction was conducted independently by I.V. and W.F., prioritizing information relevant to the study's objective.

### 2.3 Eligibility criteria

Eligible studies for this systematic review met the following criteria: (1) randomized trials evaluating efficacy or safety without time restrictions; (2) inclusion of patients with moderate to severe ulcerative colitis; (3) interventions involving selective IL-23 receptor antagonists; (4) use of placebo as the control; and (5) reporting at least one relevant outcome. Exclusion criteria were: (1) overlapping populations, defined by shared institutions and recruitment periods; (2) populations outside the scope of interest; (3) republished literature; (4) protocols without reported results; (5) reviews, case reports, case series, background articles, expert opinions, or in vivo/in vitro studies; (6) duplicate data from the same clinical trial; or (7) absence of a comparator group.

### 2.4 Outcomes measures and subgroup analysis

The effectiveness outcomes included clinical remission, clinical response, endoscopic remission, and endoscopic response during both the induction and maintenance periods. In the induction phase, additional outcomes included clinical remission in patients with previous failure of Janus kinase inhibitors or biologic therapy, histologic, endoscopic, and mucosal healing, symptomatic remission, and change in IBDQ (Inflammatory Bowel Disease Questionnaire) total score. For the maintenance phase, additional outcomes included glucocorticoid-free remission and maintenance of clinical remission. Safety outcomes encompassed the overall safety profile for both induction and maintenance periods, including any adverse events (AEs), serious AEs, and specific AEs such as anemia, headache, nasopharyngitis, and arthralgia. The endpoints' definitions are detailed in **Supplementary Table S2**. A subgroup analysis was also conducted to evaluate clinical remission outcomes during both phases, stratified by the specific type of drug administered.

### 2.5 Risk of bias

We assessed the risk of bias for RCTs using the Cochrane Risk of Bias 2 (RoB 2) tool (16). The evaluation of bias was performed by two independent reviewers (M.T. and I.V.). The RoB 2.0 tool rates the risk of bias as either high, some concerns, or low across five domains: selection, performance, detection, attrition, and reporting biases. The layout was produced by RobVis (17).

## 2.6 Certainty of evidence

The Grading of Recommendations, Assessment, Development, and Evaluation (GRADE) tool was employed by two independent authors (W.A. and M.T.) using the GRADEpro Guideline Development Tool (18) to evaluate the level of certainty of the evidence in this meta-analysis, with categorizations ranging from high to very low (19). Any disagreements were discussed and resolved through a consensus.

## 2.7 Sensitivity analysis

The stability of the pooled estimates was assessed through a leave-one-out analysis, where data from each study were sequentially removed, and the remaining dataset re-analyzed. This helped ensure that the aggregated effect sizes were not unduly influenced by any single study.

## 2.8 Statistical analysis and publication bias

Statistical analysis was performed using R software and RStudio (version 2024.04.1+748; R Core Team, Vienna, Austria), employing DerSimonian and Laird's random-effects model to calculate pooled analyses with 95% confidence intervals (CI) (20). Binary outcomes were assessed with risk ratios (RRs), continuous outcomes with mean differences (MDs), and results were displayed in forest plots. Heterogeneity was evaluated using the Cochrane Q chi-square test and I² statistic, with p-values <0.10 and I² >30% indicating significant heterogeneity (21). Publication bias was assessed with funnel plots.

### 2.9 Meta-regression analysis

To explore the impact of a history of biologic failure on clinical remission and response during induction therapy with IL-23 receptor antagonists, we conducted meta-regression analyses. The percentage of patients with a history of biologic failure was included as a covariate in the model to assess its potential influence on clinical outcomes. Clinical remission and clinical response were used as dependent variables. The analysis was performed using R software 4.4, and results were reported as estimates with 95% confidence intervals. A significance level of $p < 0.05$ was considered for statistical significance.

### 2.10 Trial sequential analysis

TSA was performed using TSA software (version 0.9.5.10 beta) (22) to assess sample size adequacy and determine the need for further research. Diversity-adjusted information size was calculated, accounting for variability between trials and sampling error, with a 5% type I error risk ($\alpha = 5\%$) and 20% type II error risk ($\beta = 20\%$) (23,24). Crossing the trial sequential monitoring boundary before reaching the required information size indicates conclusive evidence, whereas failure to cross it suggests the need for additional trials.

## 3. Results

### 3.1 Study selection

The initial search strategy yielded 1,686 results (**Fig. 1**). After the removal of duplicates and full-text screening, nine double-blind, placebo-controlled RCTs reported in six studies (11,25–29) were included in this meta-analysis. The studies reviewed during full-text screening are presented in **Supplementary Table S3**.

### 3.2 Baseline characteristics of included studies

During the induction period, six studies comprised a total of 3,808 patients, of whom 2,422 (64%) received intravenous IL-23RA (ustekinumab, mirikizumab, guselkumab, or risankizumab). The follow-up period ranged from 8 to 12 weeks. The mean age ranged from 40.5 to 42.5 years, with 2,266 (60%) being male. In addition, 1,849 (49%) of the participants had a history of failure with biologic therapies. Baseline characteristics of the induction period are detailed in **Table 1**.

Regarding the maintenance period (**Table 2**), five studies included 1,734 patients, of whom 981 (57%) received subcutaneous IL-23RA. The follow-up duration ranged from 44 to 52 weeks, and the mean age varied between 40 and 42.6 years. Among the participants, 965 (58%) were male.

### 3.3 Effectiveness and safety of the induction phase

A pooled analysis of six studies revealed that clinical remission (21.8% vs. 8%; RR 2.63; 95% CI 2.05-3.36; $p < 0.01$; $I^2 = 31\%$ [**Fig. 2A**]) and clinical response (47.5% vs. 27.1%; RR 1.94; 95% CI 1.70-2.20; $p < 0.01$; $I^2 = 39\%$ [**Fig. 2B**]) were significantly higher in the intervention group. Regarding subgroup analysis based on type of drug, there was no difference between groups for clinical remission ($p_{interaction} = 0.9$ [**Supplementary Fig. S1**]). However, among those who had a previous failure with janus-kinase inhibitors or biologic therapy, clinical remission was much higher with IL-23RA (RR 3.74; 95% CI 1.60-8.76; $p < 0.01$; $I^2 = 0\%$ [**Supplementary Fig. S2**]).

Additionally, IL-23RA were associated with significantly higher rates of endoscopic remission (RR 2.36; 95% CI 1.63-3.41; $p < 0.01$; $I^2 = 51\%$) [**Fig. 2C**]), endoscopic response (RR 2.51; 95% CI 1.99-3.16; $p < 0.01$; $I^2 = 31\%$ [**Fig. 2D**]), and histologic, endoscopic, and mucosal healing (RR 2.49; 95% CI 2.01-3.08; $p < 0.01$; $I^2 = 22\%$ [**Supplementary Fig. S3**]).

Furthermore, symptomatic remission (RR 2.16; 95% CI 1.66-2.82; p < 0.01; I² = 64% [**Supplementary Fig. S4**]) and change in the IBDQ total score (MD 16.3; 95% CI 12.2-20.4; p < 0.001; I² = 99% [**Supplementary Fig. S5**]) also significantly favored the IL-23RA.

No significant difference was observed in the rate of overall adverse events (RR 0.91; 95% CI 0.85-0.98; p = 0.01; I² = 0% [**Supplementary Fig. S6**]). However, serious adverse events were more common with placebo (RR 0.40; 95% CI 0.27-0.69; p < 0.01; I² = 25% [**Supplementary Fig. S7**]), as was the incidence of anemia (RR 0.61; 95% CI 0.46-0.82; p < 0.01; I² = 0% [**Supplementary Fig. S8**]) and worsening of ulcerative colitis (RR 0.27; 95% CI 0.18-0.38; p < 0.01; I² = 13% [**Supplementary Fig. S9**]). No differences were observed for headache (RR 1.17; 95% CI 0.82-1.68; p = 0.39; I² = 0% [**Supplementary Fig. S10**]), nasopharyngitis (RR 1.09; 95% CI 0.67-1.76; p = 0.73; I² = 0% [**Supplementary Fig. S11**]), or arthralgia (RR 1.34; 95% CI 0.77-2.32; p = 0.29; I² = 0% [**Supplementary Fig. S12**]).

### *3.4 Effectiveness and safety of the maintenance period*

A pooled analysis of five studies demonstrated a higher clinical remission (46.8% vs. 23%; RR 1.99; 95% CI 1.63-2.44; p < 0.01; I² = 38% [**Fig. 3A**]) and clinical response (75.4% vs. 48.7%; RR 1.51; 95% CI 1.34-1.69; p < 0.01; I² = 21% [**Fig. 3B**]) with IL-23RA. Subgroup analysis showed no difference regarding type of drug for clinical remission (p_{interaction} = 0.15 [**Supplementary Fig. S13**]).

IL-23RA demonstrated superior efficacy in achieving endoscopic remission (RR: 1.96; 95% CI: 1.63–2.37; *p* < 0.01; I² = 0% [**Fig. 3C**]) and endoscopic response (RR: 2.20; 95% CI: 1.51–2.70; *p* < 0.01; I² = 61% [**Fig. 3D**]). Furthermore, IL-23RA significantly improved glucocorticoid-free remission rates (RR: 1.98; 95% CI: 1.60–2.45; *p* < 0.01; I² = 46% [**Supplementary Fig. S14**]) and was associated with a higher maintenance of clinical remission (RR: 1.80; 95% CI: 1.48–2.18; *p* < 0.01; I² = 0% [**Supplementary Fig. S15**]).

The safety profile was comparable between groups. There was no difference regarding the overall incidence of adverse events (RR 0.97; 95% CI 0.92-1.03; p = 0.32; I² = 0% [**Supplementary Fig. S16**]), nasopharyngitis (RR 1.07; 95% CI 0.77-1.50; p = 0.68; I² = 0% [**Supplementary Fig. S17**]), arthralgia (RR 1.10; 95% CI 0.72-1.70; p = 0.66; I² = 15% [**Supplementary Fig. S18**]), headache (RR 1.29; 95% CI 0.60-2.78; p = 0.51; I² = 61%) [**Supplementary Fig. S19**]), serious adverse events (RR 0.75; 95% CI 0.31-1.84; p = 0.53; I² = 64%) [**Supplementary Fig. S20**]), or anemia (RR 0.62; 95% CI 0.35-1.07; p = 0.09; I² = 0% [**Supplementary Fig. S21**]). However, worsening of ulcerative colitis (RR 0.41; 95% CI 0.25-0.66; p < 0.01; I² = 73% [**Supplementary Fig. S22**]) was higher in the placebo group.

### 3.5 Risk of bias within studies

As presented in **Supplementary Fig. S23**, six RCTs were assessed as having a low risk of bias (11,25–27,29). Conversely, three studies were categorized as having some concerns of bias, mainly due to issues related to the randomization process (11,28) or deviations from intended intervention (26).

### 3.6 Certainty of evidence and publication bias

According to the GRADE criteria (**Supplementary Table S4**), the certainty of evidence was high for the outcomes of clinical remission/response and endoscopic remission/response during the induction phase. In the maintenance phase, certainty of evidence was moderate for clinical remission/response and low for endoscopic remission/response. Funnel plot analysis (**Supplementary Figs. S24-S27**) showed no indications of publication bias, with symmetrical plots observed for outcomes.

### 3.7 Sensitivity analysis

We performed leave-one-out sensitivity analysis to assess the influence of individual studies on the pooled results. When any single study was omitted, the p-values did not cross the threshold for statistical significance, indicating that the results were robust and not sensitive to the exclusion of individual studies. Leave-one-out sensitivity analyses are detailed in **Supplementary Figs. S28-S31**.

### 3.8 Meta-regression analysis

The analysis (**Supplementary Fig. S32**) identified a significant association between a history of biologic failure and clinical remission (p = 0.008) but not with clinical response (p = 0.59). Minimal residual heterogeneity was observed, with tau² = 0 and I² = 0% for clinical remission, indicating that moderators explained nearly all variability. The model's R² value of 100% highlighted its effectiveness in accounting for differences in clinical remission outcomes across studies.

### 3.9 Trial sequential analysis

The required information size (RIS), calculated to target a 5% risk of type I error and a 20% risk of type II error, was 345 patients for clinical remission in the induction phase (**Supplementary Fig. S33**) and 229 patients for the maintenance phase (**Supplementary Fig. S34**). For clinical response, the RIS was 336 patients in the induction phase (**Supplementary Fig.. S35**) and 144 patients in the maintenance phase (**Supplementary Fig. S36**). The cumulative Z-curve crossed the monitoring boundary for all endpoints, indicating that the number of included studies was adequate to provide robust evidence.

## 4. Discussion

In this systematic review and meta-analysis of 9 RCTs involving 3,808 patients, we evaluated the effectiveness and safety of IL-23RA in the induction and maintenance treatment phases for UC. The main findings from the pooled analysis were: (1) IL-23RA significantly improved clinical remission, clinical response, and endoscopic remission compared to placebo in the induction phase; (2) similar benefits were observed during the maintenance phase, with IL-23RA continuing to show superior clinical and endoscopic outcomes; and (3) IL-23RA demonstrated a comparable safety profile to placebo, with a lower incidence of disease worsening.

To our knowledge, this is the first pooled analysis to comprehensively assess the efficacy and safety of selective IL-23RA in patients with moderate to severe UC. While other systematic reviews and NMA have evaluated biologic therapies and small molecules for UC, they are limited in scope, often focusing solely on efficacy outcomes and lacking direct comparisons of IL-23RA as a class (12,13,30,31). Therefore, this meta-analysis provides a more focused assessment of the IL-23 receptor antagonists class, incorporating the latest updated data and employing robust methodological approaches.

The effectiveness of IL-23RA in UC may be attributed to their mechanism of action, targeting the IL-23/Th17 pathway, which plays a critical role in the pathogenesis of UC (7). IL-23 is involved in the differentiation and maintenance of Th17 cells, which are implicated in driving the inflammatory process in UC (7,9). By inhibiting the interaction between IL-23 and its receptor, these antagonists may effectively reduce inflammation and mucosal damage, leading to both clinical improvement and endoscopic healing (9,32). Specific IL-23R antagonists include ustekinumab, a fully human monoclonal antibody that targets both IL-23 and IL-12; mirikizumab, a monoclonal antibody targeting the p19 subunit of IL-23, shown to reduce inflammation and improve symptoms; guselkumab, another monoclonal antibody that selectively targets the p19 subunit of IL-23, primarily studied in psoriasis; and risankizumab, which binds to the p19 subunit of IL-23 and has demonstrated promising results in UC trials, with potential for inducing long-term remission (33). These agents, by selectively modulating the IL-23/Th17 axis, provide targeted therapeutic options for managing UC.

This meta-analysis demonstrates that IL-23RA are highly effective in both inducing and maintaining clinical remission in patients with UC. During the induction phase, pooled analyses showed significantly higher rates of clinical remission, clinical response, and endoscopic remission in the IL-23RA group compared to placebo. In the maintenance phase, these benefits were sustained, with increased rates of clinical remission and response, as well as continued endoscopic improvements, supporting the long-term efficacy of IL-23RA. These findings align with current treatment goals in UC (34–36), which emphasize achieving not only symptomatic relief but also durable endoscopic and clinical outcomes, further underscoring the value of IL-23RA in comprehensive disease management.

In our meta-analysis, clinical remission during both the induction and maintenance phases was not significantly affected by any specific IL-23RA, as subgroup analysis showed no statistical significant differences between the drugs. This contrasts with a recent NMA that ranked risankizumab second in clinical remission, followed by guselkumab, and highlighted their strong performance in histological remission induction (12). While drugs like risankizumab and guselkumab may excel in certain contexts, our findings suggest that the clinical remission achieved with IL-23RAs is likely a class effect, rather than driven by a single agent, challenging the NMA's emphasis on individual drug efficacy.

We performed meta-regression analyses to assess how a history of biologic failure affects clinical remission and response during induction therapy with IL-23RA. The results showed a significant link between biologic failure and clinical remission, indicating that populations with a higher rate of biologic failure may achieve clinical remission more effectively. Additionally, we found that clinical remission rates were considerably higher with IL-23RA among patients with a history of biologic failure. However, we did not find a significant association for clinical response. One possible explanation is that IL-23RA may be more effective in achieving stringent outcomes like remission, while response, as a broader measure, may be less affected by prior biologic therapy failure. These findings contrast with existing literature, which suggests that patients with a history of biologic failure typically experience diminished responses to treatment due to factors like uncontrolled inflammation and structural or immunological changes in the colon (37). Our results should be interpreted cautiously, as they highlight the complexity of treatment outcomes and the need for further research to reconcile these findings and refine therapeutic strategies.

Additionally, this study highlights significant improvements in both histologic and symptomatic relief in patients with UC. While the primary goals of UC treatment should focus on endoscopic and clinical improvements, histologic remission is regarded as a marker of deep remission, which plays a crucial role in achieving long-term disease control (38). Additionally, patients experienced significant symptomatic relief, as measured by the IBDQ score, which evaluates bowel, social, emotional, and systemic functions that impact quality of life (39). However, the interpretation of the IBDQ score improvements should be approached with caution due to the high heterogeneity observed ($I^2 = 99\%$), indicating variability in patient responses across studies.

This meta-analysis underscores the favorable safety profile of IL-23RA in ulcerative colitis. During induction, these agents significantly reduced serious AE, disease worsening, and anemia, without increasing overall adverse events. In the maintenance phase, IL-23RA lowered the incidence of disease worsening while demonstrating comparable rates of AE and anemia. Notably, IL-23RA did not elevate risks of common adverse events such as headache, nasopharyngitis, or arthralgia, highlighting their safety compared to other therapeutic options. Traditional treatments like corticosteroids and immunomodulators are first-line therapies, but with limited efficacy (1,6). Although advanced therapies such as TNF-α inhibitors and Janus kinase inhibitors (JAKi) are effective, their use is often restricted due to concerns about severe infections and thromboembolic events, especially with JAKi (40,41). These findings reinforce the role of IL-23RA as a safe and more targeted therapeutic option for patients with moderate to severe UC.

This meta-analysis shows consistent effectiveness of IL-23RA in both the induction and maintenance phases. Long-term extension data from the LUCENT-3 study (42) with mirikizumab (up to week 152) and the UNIFI study (43) with ustekinumab (up to week 200) further support sustained efficacy, with significant clinical response and endoscopic improvement over time. These findings highlight the ability of IL-23RA to provide long-term therapeutic benefits in ulcerative colitis.

Additionally, IL-23RA significantly improved glucocorticoid-free remission rates and enhanced the maintenance of clinical remission. Achieving glucocorticoid-free remission is particularly significant, as it reduces the risks associated with long-term corticosteroid use, such as weight gain, diabetes, osteoporosis, bone damage, and a higher risk of infections (44). The continued improvement in clinical remission further highlights the long-term benefits of IL-23RA in reducing steroid dependence and optimizing disease management.

The TSA in this meta-analysis confirms the robustness of the findings on clinical remission/response during both induction and maintenance phases. The Z-curve crossing the monitoring boundary indicates substantial evidence supporting the efficacy of IL-23RA, strengthening confidence in their effectiveness for inducing and maintaining clinical remission and response in UC.

Future research should focus on identifying patient populations most likely to benefit from IL-23RA by stratifying them based on biomarkers, disease severity, and prior treatments. This would help optimize treatment selection and improve outcomes. Additionally, head-to-head trials comparing IL-23RA with other biologics could provide valuable insights into their relative                    efficacy                    and                    safety.

This study has several limitations that warrant consideration. First, there was slight variability in the definitions of primary outcomes across the included studies, which may influence the consistency of the results. To mitigate this, we performed sensitivity analyses to ensure that the conclusions were not disproportionately affected by any single study. Second, some degree of heterogeneity was observed in certain outcomes. However, this was accounted for by employing a random-effects model to adjust for the variability observed across the studies. Finally, although the analysis included a relatively small number of studies, the TSA demonstrated that the sample size was sufficient to support the robustness of the findings.

## 5. Conclusion

Selective IL-23RA are a promising new treatment for moderate to severe UC, showing superior effectiveness over placebo in achieving clinical, endoscopic, histologic and symptomatic improvements with a favorable safety profile. These findings highlight the potential of IL-23RA as a key therapeutic target for UC, offering an effective and safe option for managing this challenging condition.

Table 1. Baseline characteristics of the induction period.

| Study | Year | Phase | Follow-up (Weeks) | IL-23RA (dosage) | Sample size | Male (%) | Age, year (SD) | Weight, kg (SD) | Disease duration, years (SD) | Total Mayo score (SD) | Severe UC †, n (%) | Failure with biologics, n (%) | C-reactive protein, mg/liter (SD) | Fecal calprotectin, mg/kg (SD)(29) |
|---|---|---|---|---|---|---|---|---|---|---|---|---|---|---|
| UNIFI (29) | 2019 | 3 | 8 | Ustekinumab 130 mg (single dose) | 639 | 387 (60.6) | 41.7 (13.7) | 73.3 (18.4) | 8.05 (7.2) | 8.9 (1.6)§ | NA | 352 (50.9) | 4.6 (1.5) | 1,313 (344) |
| Sandborn (28) | 2020 | 2 | 12 | Mirikizumab 200 mg q4W | 125 | 73 (58.3) | 43 (14) | 74.8 (16.7) | 9.25 (9.3) | 6.6 (1.3) * | 70 (56) | 74 (59.2) | 4.22 (2.4) | 1,460 (425) |
| QUASAR Phase 3 (25) | 2024 | 3 | 12 | Guselkumab 200 mg q4W | 701 | 399 (56.9) | 40.5 (13.7) | 72.5 (16.8) | 7.5 | 6.9 (1.1) | 452 (68.5) | 344 (49) | 5.4 (6.6) | 1878 (1991) |
| QUASAR Phase 2b (27) | 2019 | 2b | 12 | Guselkumab 200 mg q4W | 206 | 126 (61.2) | 42.2 (14.7) | 69.6 (16.4) | 7.35 (6.6) | 6.9 (1.6)* | 140 (68) | 97 (47.1) | 5 (2.6) | 1,595 (453) |
| INSPIRE (26) | 2024 | 3 | 12 | Risankizumab 1200 mg q4W | 975 | 586 (60.1) | 42.1 (13.7) | NA | 7.9 (6.9) | 7.1 (1.2)* | 408 (42) | 503 (51.6) | 3.7 (1.3) | 1,576 (454) |
| LUCENT-1 (11) | 2024 | 3 | 12 | Mirikizumab 300 mg q4W | 1,162 | 695 (59.8) | 42.5 (13.7) | NA | 7.05 (6.85) | NA | 618 (53) | 479 (41.2) | 4.2 (1.3) | 1,548 (406) |

Footnotes: Continuous variables are presented as mean at baseline (Standard Deviation, SD). NA, not available; SD, standard deviation; *Modified Mayo score; †Modified Mayo score 7–9; §Mayo score; q4W, every four weeks.

Table 2. Baseline characteristics of the maintenance period.

| Study | Year | Follow-up (Weeks) | IL-23RA | Sample size | Male (%) | Age, years (SD) | Total Mayo score (SD) | C-reactive protein, mg/liter (SD) | Fecal calprotectin, mg/kg (SD) |
|---|---|---|---|---|---|---|---|---|---|
| UNIFI (29) | 2019 | 44 | Ustekinumab 90 mg q8w | 351 | 201 (57.3) | 40.7 (13.6) | 8.8 (1.5)§ | 3.9 (1.9) | 1,511 (402) |
| Sandborn 2020 (28) | 2020 | 52 | Mirikizumab 200 mg q4W | 60 | NA | NA | NA | NA | NA |
| QUASAR Phase 3 (25) | 2024 | 44 | Guselkumab 200 mg q4W | 380 | 209 (55) | 40.9 (14.1) | 6.95 (1)* | 4.8 (5.4) | 1,834 (1,971) |
| COMMAND (26) | 2024 | 52 | Risankizumab 180 mg q8w | 362 | 212 (58.6) | 40 (14.5) | 7.2 (1.2)* | 4.7 (2.1) | 1,584 (451) |
| LUCENT-2 (11) | 2023 | 40 | Mirikizumab 200 mg q4W | 581 | 343 (59) | 42.6 (13.8) | NA | NA | NA |

Footnotes: Continuous variables are presented as mean at baseline (Standard Deviation, SD). NA, not available; SD, standard deviation; *Modified Mayo score; §Mayo score; q4W, every four weeks; q8W, every eight weeks.

10 **Fig. 1**. Preferred Reporting Items for Systematic Reviews and Meta-Analysis (PRISMA) flow diagram of study screening and selection.

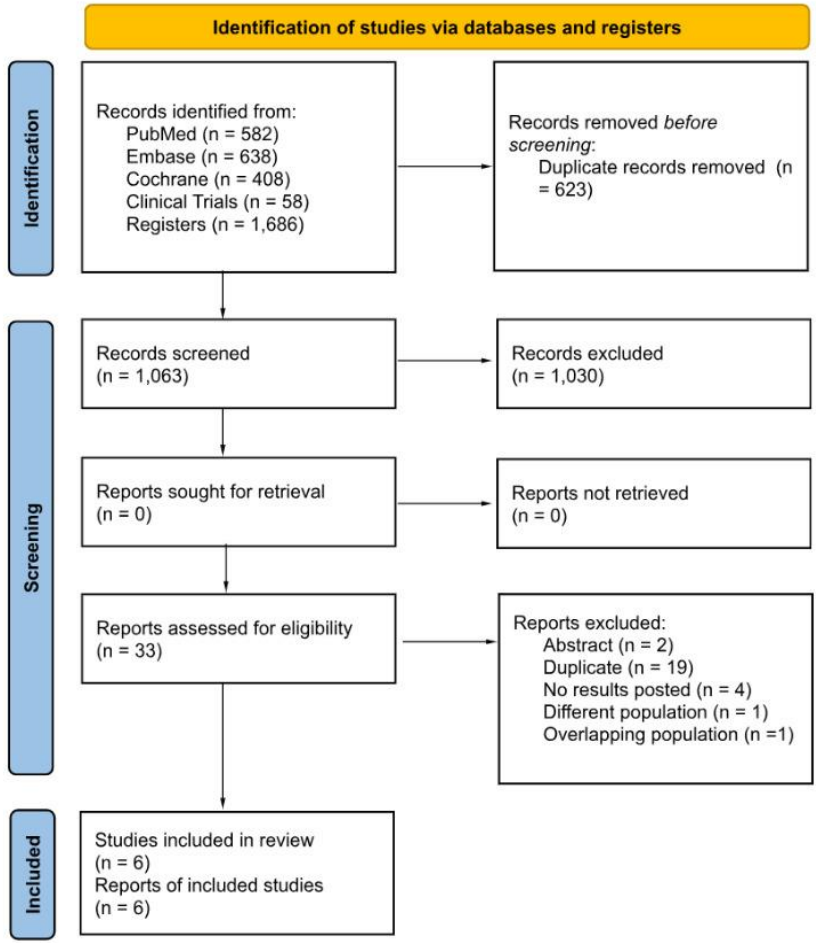

14 **Fig. 2**. Forest plots for (A) clinical remission, (B) clinical response, (C) endoscopic remission, and (D) endoscopic response in the induction phase.

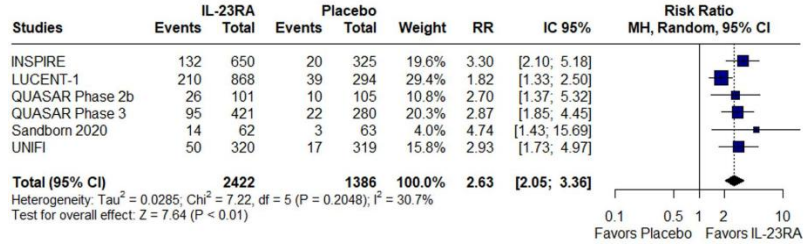

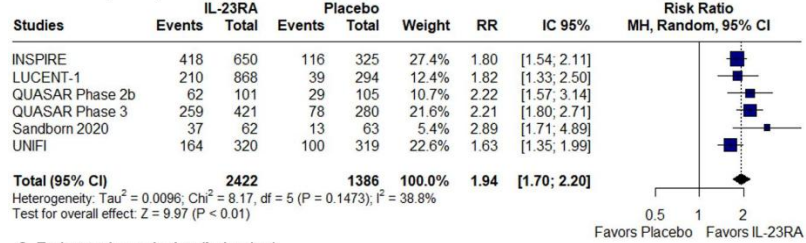

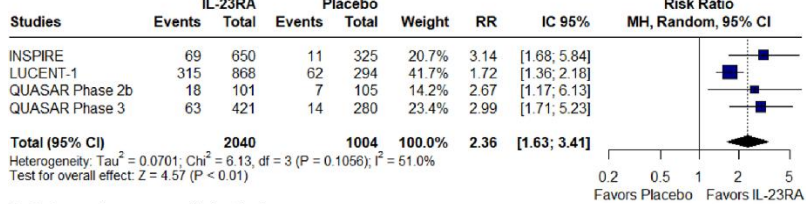

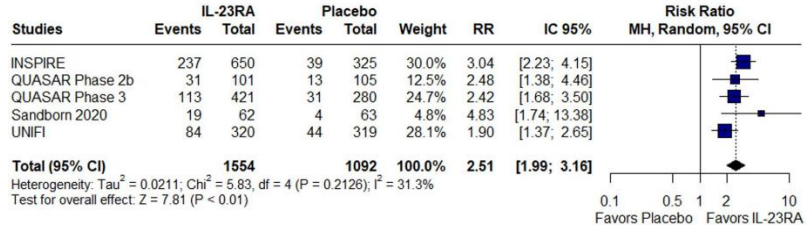

17  **Fig. 3**. Forest plots for (A) clinical remission, (B) clinical response, (C) endoscopic remission, and (D) endoscopic response in the maintenance

18  phase.

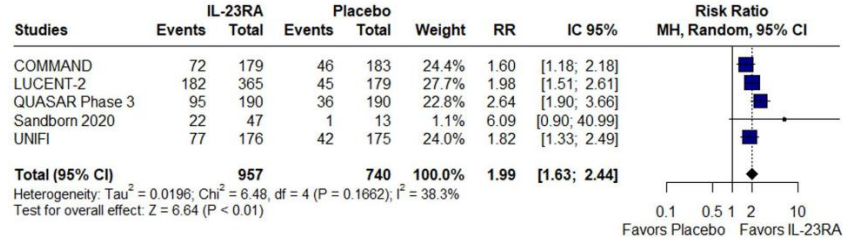

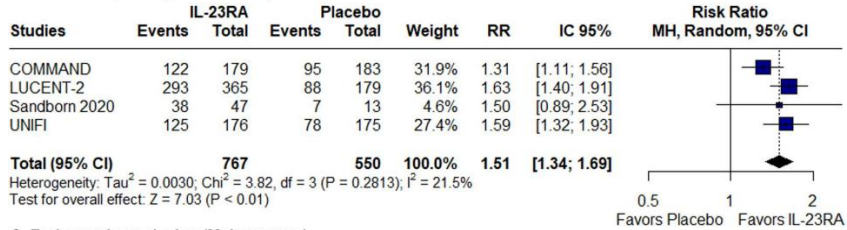

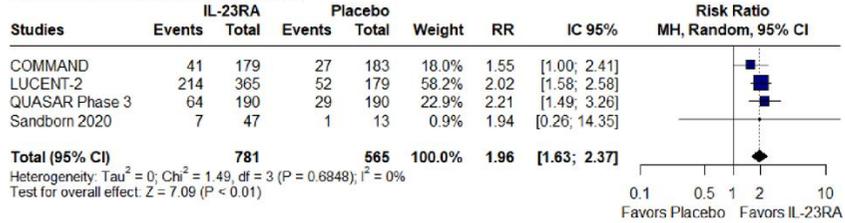

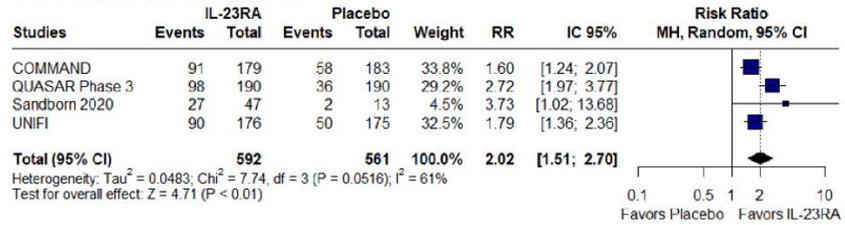

*SUPPLEMENTARY MATERIAL*

**Effectiveness and Safety of Selective IL-23 Receptor Antagonists in Moderate to Severe Ulcerative Colitis: A Systematic Review, Meta-analysis, and Trial Sequential Analysis**

**Supplementary Table S1**. Search strategy for each database.

| PubMed |
| --- |
| ("ulcerative colitis" OR ulcerative colitis[MeSH Terms]) AND ("IL-23" OR "Interleukin-23" OR Guselkumab OR Tildrakizumab OR Mirikizumab OR Ustekinumab OR Risankizumab OR IL-23[MeSH Terms] OR Interleukin-23[MeSH Terms] OR Ustekinumab[MeSH Terms]) AND ("randomized controlled trial"[pt] OR "controlled clinical trial"[pt] OR randomized[tiab] OR placebo[tiab] OR "drug therapy"[sh] OR randomly[tiab] OR trial[tiab] OR groups[tiab]) |
| **Cochrane** |
| ("ulcerative colitis") AND ("IL-23" OR "Interleukin-23" OR Guselkumab OR Tildrakizumab OR Mirikizumab OR Ustekinumab OR Risankizumab) AND ("randomized controlled trial" OR "controlled clinical trial" OR randomized OR placebo OR "drug therapy" OR randomly OR trial OR groups) |
| **Embase** |
| ('ulcerative colitis' OR 'ulcerative colitis'/exp) AND ('il-23' OR 'interleukin-23' OR 'guselkumab' OR 'tildrakizumab' OR 'mirikizumab' OR 'ustekinumab' OR 'risankizumab' OR 'interleukin 23'/exp OR 'ustekinumab'/exp) AND ('randomized controlled trial':it OR 'controlled clinical trial':it OR 'randomized':ti,ab,kw OR 'placebo':ti,ab,kw  OR 'randomly':ti,ab,kw) |
| **Clinical trials** |
| ("ulcerative colitis") AND ("IL-23" OR "Interleukin-23" OR Guselkumab OR Tildrakizumab OR Mirikizumab OR Ustekinumab OR Risankizumab) AND ("randomized controlled trial" OR "controlled clinical trial" OR randomized OR placebo OR "drug therapy" OR randomly OR trial OR groups) |

**Supplementary Table S2**. Endpoints' definition.

| Study | Outcome | Definition |
| --- | --- | --- |
| Louis 2024 (INSPIRE, COMMAND) | Clinical Remission (Induction, Week 12) | Stool frequency ≤1, rectal bleeding score of 0, endoscopic subscore ≤1 without friability. |
| | Clinical Remission (Maintenance, Week 52) | Same criteria as for induction trial using the adapted Mayo score. |
| | Clinical Response | ≥30% and ≥2-point decrease in adapted Mayo score, rectal bleeding score decrease ≥1 or ≤1. |
| | Endoscopic Improvement | Endoscopic subscore ≤1 without friability. |
| | Endoscopic Remission | Endoscopic subscore of 0. |
| | Histological and Mucosal Improvement | Endoscopic subscore ≤1, Geboes score ≤3.1. |
| | Histological and Mucosal Remission | Endoscopic subscore of 0, Geboes score <2.0. |
| Sands 2019 (UNIFI) | Clinical Remission | Total Mayo score ≤2, no subscore >1. |
| | Endoscopic Improvement | Mayo endoscopic subscore of 0 or 1. |
| | Clinical Response | ≥30% and ≥3-point reduction in Mayo score with rectal bleeding subscore decrease ≥1 or ≤1. |
| | Histo-Endoscopic Mucosal Healing | Histologic improvement (neutrophil infiltration <5%, no crypt destruction or erosion) and endoscopic improvement. |
| D'Haens 2023 (LUCENT-1, LUCENT-2) | Clinical Remission (Induction, Week 12) | Stool-frequency subscore of 0 or 1 with ≥1-point decrease from baseline, rectal-bleeding subscore of 0, and endoscopic subscore of 0 or 1 (excluding friability). |
| | Clinical Remission (Maintenance, Week 40) | Same criteria as for induction trial. |
| | Clinical Response | ≥2-point and ≥30% decrease from baseline in modified Mayo score, plus rectal-bleeding subscore of 0 or 1 or ≥1-point decrease from baseline. |
| | Endoscopic Remission | Endoscopic subscore of 0 or 1 (excluding friability). |
| | Remission of Symptoms | Stool-frequency subscore of 0 or 1 with ≥1-point decrease from baseline, and rectal-bleeding subscore of 0. |
| | Histologic–Endoscopic Mucosal Improvement | Endoscopic remission and histologic improvement (neutrophil infiltration in <5% of crypts, no crypt destruction, erosions, ulcerations, or granulation tissue). |
| Peyrin-Biroulet 2023 (QUASAR Phase 2b) | Clinical Response (Week 12) | Decrease in modified Mayo score ≥30% and ≥2 points, with either a ≥1-point decrease in the rectal bleeding subscore or a rectal bleeding subscore of 0 or 1. |
| | Clinical Remission | Mayo stool frequency subscore of 0 or 1 and not increased from baseline, rectal bleeding subscore of 0, and endoscopy subscore of 0 or 1 with no friability. |
| | Symptomatic Remission | Mayo stool frequency subscore of 0 or 1 and not increased from baseline, and rectal bleeding subscore of 0. |
| | Endoscopic Improvement | Mayo endoscopy subscore of 0 or 1 with no friability. |
| | Histo-Endoscopic Mucosal Improvement | Endoscopic improvement plus histologic improvement (neutrophil infiltration <5% of crypts, no crypt destruction, erosions, ulcerations, or granulation tissue). |
| Sandborn 2020 | Clinical Remission (Week 12) | Mayo rectal bleeding subscore of 0, stool frequency subscore of 0 or 1 with a ≥1-point decrease from baseline, and endoscopy subscore of 0 or 1. |
| | Clinical Response (Weeks 12 and 52) | Decrease in Mayo subscore ≥2 points and ≥35% from baseline, with either a rectal bleeding subscore decrease of ≥1 or a rectal bleeding subscore of 0 or 1. |
| | Durable Clinical Remission | Clinical remission achieved at both Weeks 12 and 52. |
| | Endoscopic Remission | Mayo endoscopy subscore of 0. |
| | Endoscopic Improvement | Mayo endoscopy subscore of 0 or 1. |
| | Histologic Remission | Geboes histologic subscores of 0 for neutrophils in lamina propria, neutrophils in epithelium, and erosion or ulceration parameters. |
| | Change in IBDQ Score | Change from baseline in Inflammatory Bowel Disease Questionnaire (IBDQ). |
| | Symptomatic Remission | Stool frequency subscore of 0 or 1 and rectal bleeding subscore of 0. |
| **Phase 3 QUASAR Maintenance Study** | Clinical Remission (Week 44) | Mayo stool frequency subscore of 0 or 1 with a ≥1-point decrease from baseline, rectal bleeding subscore of 0, and Mayo endoscopic subscore of 0 or 1. |
| | Maintenance of Clinical Response | Sustained clinical response through Week 44. |
| | Symptomatic Remission | Stool frequency subscore of 0 or 1 and rectal bleeding subscore of 0. |
| | Endoscopic Remission | Mayo endoscopic subscore of 0. |
| | Endoscopic Improvement | Mayo endoscopic subscore of 0 or 1. |
| | Histologic-Endoscopic Mucosal Improvement | Combined endoscopic improvement and histologic improvement (e.g., reduced neutrophil infiltration and no erosions or ulcerations). |
| **Phase 3 QUASAR Induction Study** | Clinical Remission (Week 12) | Mayo stool frequency subscore of 0 or 1 with no increase from baseline, rectal bleeding subscore of 0, and Mayo endoscopy subscore of 0 or 1 without friability. |
| | Symptomatic Remission (Week 12) | Stool frequency subscore of 0 or 1 with no increase from baseline, and rectal bleeding subscore of 0. |
| | Clinical Response (Week 12) | ≥30% and ≥2-point reduction in modified Mayo score, with either a rectal bleeding subscore decrease of ≥1 or a rectal bleeding subscore of 0 or 1. |
| | Endoscopic Improvement (Week 12) | Mayo endoscopy subscore of 0 or 1. |
| | Histologic-Endoscopic Improvement (Week 12) | Combined endoscopic improvement and histologic improvement (e.g., reduced neutrophil infiltration, no crypt destruction, no erosions). |

**Supplementary Table S3**. Table of the studies included in the full-text screening.

| | Title | Author/year | Final decision | Reason for exclusion |
|---|---|---|---|---|
| 1 | A Study to Assess the Efficacy and Safety of Risankizumab in Participants With Ulcerative Colitis | Clinical trials (NCT03398135) | Excluded | Duplicate |
| 2 | A MAINTENANCE STUDY OF THE EFFICACY AND SAFETY OF MIRIKIZUMAB IN ULCERATIVE COLITIS | EUCTR2017-003238-96-ES | Excluded | No results posted |
| 3 | A Study of Mirikizumab (LY3074828) in Participants With Moderate to Severe Ulcerative Colitis | Clinical trials (NCT02589665) | Excluded | Duplicate |
| 4 | Efficacy, Safety, and Pharmacokinetics of Guselkumab in Pediatric Participants with Moderately to Severely Active Ulcerative Colitis | Clinical trials | Excluded | Duplicate |
| 5 | A Study of Guselkumab Therapy in Participants With Moderately to Severely Active Ulcerative Colitis | Clinical trials (NCT02589665) | Excluded | Duplicate |
| 6 | A Study of the Efficacy and Safety of Guselkumab in Participants with Moderately to Severely Active Ulcerative Colitis | Clinical trials (NCT05528510) | Excluded | No results posted |
| 7 | A Study to Evaluate the Safety and Efficacy of the Ustekinumab Induction and Maintenance Therapy in Participants with Moderately to Severely Active Ulcerative Colitis | Clinical trials | Excluded | No results posted |
| 8 | A Phase 3, Multicenter, Randomized, Double-Blind, Parallel, Placebo-Controlled Induction Study of Mirikizumab in Conventional-Failed and Biologic-Failed Patients With Moderately to Severely Active Ulcerative Colitis (LUCENT 1) | Clinical trials | Excluded | Duplicate |
| 9 | A Phase 3, Multicenter, Randomized, Double-Blind, Parallel-Arm, Placebo-Controlled Maintenance Study of Mirikizumab in Patients With Moderately to Severely Active Ulcerative Colitis (LUCENT 2) | Clinical trials | Excluded | Duplicate |
| 10 | A Multicenter, Randomized, Double-Blind, Placebo Controlled 52-Week Maintenance and an Open-Label Extension Study of the Efficacy and Safety of Risankizumab in Subjects With Ulcerative Colitis | Clinical trials | Excluded | Duplicate |
| 11 | A Phase 3, Randomized, Double-blind, Placebo-controlled, Parallel-group, Multicenter Study to Evaluate the Efficacy and Safety of Guselkumab Subcutaneous Induction Therapy in Participants With Moderately to Severely Active Ulcerative Colitis | Clinical trials | Excluded | Duplicate |
| 12 | A Phase 3 Randomized, Open-label Induction, Double-blind Maintenance, Parallel-group, Multicenter Protocol to Evaluate the Efficacy, Safety, and Pharmacokinetics of Guselkumab in Pediatric Participants With Moderately to Severely Active Ulcerative Colitis | Clinical trials | Excluded | No results posted |
| 13 | A Phase 3 Study of the Efficacy, Safety and Pharmacokinetics of Ustekinumab as Open-label Intravenous Induction Treatment Followed by Randomized Double-blind Subcutaneous Ustekinumab Maintenance in Pediatric Participants With Moderately to Severely Active Ulcerative Colitis | Clinical trials | Excluded | Duplicate |
| 14 | A Phase 3, Randomized, Double-blind, Placebo-controlled, Parallel-group, Multicenter Protocol to Evaluate the Safety and Efficacy of Ustekinumab Induction and Maintenance Therapy in Subjects With Moderately to Severely Active Ulcerative Colitis | Clinical trials | Excluded | Duplicate |
| 15 | Risankizumab for Ulcerative Colitis: Two Randomized Clinical Trials. | Louis 2024 | Included | - |
| 16 | One-year outcomes with ustekinumab therapy in infliximab-refractory paediatric ulcerative colitis: a multicentre prospective study. | Dhaliwal J 2021 | Excluded | Duplicate |
| 17 | Three-Year Efficacy and Safety of Mirikizumab Following 152 Weeks of Continuous Treatment for Ulcerative Colitis: Results From the LUCENT-3 Open-Label Extension Study. | Sands BE 2024 | Excluded | Overlapping population |
| 18 | The Efficacy and Safety of Guselkumab Induction Therapy in Patients With Moderately to Severely Active Ulcerative Colitis: Results From the Phase 3 QUASAR Induction Study | Abstract | Excluded | Abstract |
| 19 | GUSELKUMAB (GUS) ON HISTOLOGIC AND ENDOSCOPIC OUTCOMES IN MODERATELY TO SEVERELY ACTIVE ULCERATIVE COLITIS: RESULTS FROM PHASE 3 QUASAR STUDY | Huang, K 2024 | Excluded | Duplicate |
| 20 | 984 RISANKIZUMAB MAINTENANCE THERAPY IN PATIENTS WITH MODERATELY TO SEVERELY ACTIVE ULCERATIVE COLITIS: EFFICACY AND SAFETY IN THE RANDOMIZED PHASE 3 COMMAND STUDY | Schreiber, S 2024 | Excluded | Duplicate |
| 21 | GUSELKUMAB IMPROVES HEALTH-RELATED QUALITY OF LIFE AS MEASURED BY PROMIS-29 IN PATIENTS WITH MODERATELY TO SEVERELY ACTIVE ULCERATIVE COLITIS: PHASE 3 QUASAR INDUCTION STUDY | Panés, J. 2024 | Excluded | Duplicate |
| 22 | 904 ADDITIONAL RISANKIZUMAB THERAPY IS EFFECTIVE IN PATIENTS WITH MODERATELY TO SEVERELY ACTIVE ULCERATIVE COLITIS WHO DID NOT ACHIEVE CLINICAL RESPONSE TO INITIAL 12-WEEK INDUCTION THERAPY: AN ANALYSIS OF PHASE 3 INSPIRE AND COMMAND STUDIES | Panaccione 2024 | Excluded | Duplicate |
| 23 | GUSELKUMAB IMPROVES SYMPTOMS OF FATIGUE IN PATIENTS WITH MODERATELY TO SEVERELY ACTIVE ULCERATIVE COLITIS: PHASE 3 QUASAR INDUCTION STUDY RESULTS AT WEEK 12 | Dignass, A 2024 | Excluded | Duplicate |
| 24 | MIRIKIZUMAB IMPROVES FATIGUE, BOWEL URGENCY, AND QUALITY OF LIFE IN PATIENTS WITH MODERATELY TO SEVERELY ACTIVE CROHN'S DISEASE: RESULTS FROM A PHASE 3 CLINICAL TRIAL | Ghosh, S 2024 | Excluded | Different population |
| 25 | EFFICACY OUTCOMES OF PLACEBO MAINTENANCE TREATMENT IN PATIENTS WITH MODERATELY TO SEVERELY ACTIVE ULCERATIVE COLITIS WHO RESPONDED TO PLACEBO INDUCTION THERAPY: RESULTS FROM THE PHASE 3 COMMAND STUDY | Atreya, R 2024 | Excluded | Duplicate |
| 26 | Guselkumab in Patients With Moderately to Severely Active Ulcerative Colitis: QUASAR Phase 2b Induction Study | Peyrin-Biroulet, L 2023 | Included | - |
| 27 | Risankizumab Induction Therapy in Patients With Moderately to Severely Active Ulcerative Colitis: Efficacy and Safety in the Randomized Phase 3 INSPIRE Study | Louis, E 2023 | Excluded | Duplicate |
| 28 | Mirikizumab as Induction and Maintenance Therapy for Ulcerative Colitis | D'Haens 2023 | Included | - |

| 29 | The efficacy and safety of guselkumab induction therapy in patients with moderately to severely active Ulcerative Colitis: Phase 2b QUASAR Study results through week 12 | Dignass, A. 2022 | Excluded | Duplicate |
|---|---|---|---|---|
| 30 | The Efficacy and Safety of Guselkumab as Maintenance Therapy in Patients With Moderately to Severely Active Ulcerative Colitis: Results From the Phase 3 QUASAR Maintenance Study | Abstract | Excluded | Abstract |
| 31 | Efficacy and Safety of Mirikizumab in a Randomized Phase 2 Study of Patients With Ulcerative Colitis | Sandborn 2020 | Included | - |
| 32 | Ustekinumab as Induction and Maintenance Therapy for Ulcerative Colitis | Sands 2019 | Included | - |
| 33 | Guselkumab in patients with moderately to severely active ulcerative colitis (QUASAR): phase 3 double-blind, randomised, placebo-controlled induction and maintenance studies | Rubin 2024 | Included | - |

**Supplementary Table S4**. GRADE assessment.

| Outcomes | № of participants (studies) | Certainty of the evidence (GRADE) | Relative effect (95% CI) |
|---|---|---|---|
| Clinical remission (induction) | 3808 (6 RCTs) | ⊕⊕⊕⊕ High | RR 2.63 (2.05 to 3.36) |
| Clinical response (induction) | 3808 (6 RCTs) | ⊕⊕⊕⊕ High | RR 1.94 (1.70 to 2.20) |
| Endoscopic remission (induction) | 2343 (3 RCTs) | ⊕⊕⊕⊕ High | RR 2.36 (1.70 to 2.20) |
| Endoscopic response (induction) | 2646 (5 RCTs) | ⊕⊕⊕⊕ High | RR 2.51 (1.99 to 3.16) |
| Clinical remission (maintenance) | 1697 (5 RCTs) | ⊕⊕⊕◯ Moderate[a] | RR 1.99 (1.63 to 2.44) |
| Clinical response (maintenance) | 1317 (4 RCTs) | ⊕⊕⊕◯ Moderate[a] | RR 1.51 (1.34 to 1.69) |
| Endoscopic remission (maintenance) | 966 (3 RCTs) | ⊕⊕◯◯ Low[a,b] | RR 1.96 (1.63 to 2.37) |
| Endoscopic response (maintenance) | 773 (3 RCTs) | ⊕⊕◯◯ Low[a,b] | RR 2.02 (1.51 to 2.70) |

\***The risk in the intervention group** (and its 95% confidence interval) is based on the assumed risk in the comparison group and the **relative effect** of the intervention (and its 95% CI).

**CI:** confidence interval; **RR:** risk ratio

**GRADE Working Group grades of evidence**

**High certainty:** we are very confident that the true effect lies close to that of the estimate of the effect.

**Moderate certainty:** we are moderately confident in the effect estimate: the true effect is likely to be close to the estimate of the effect, but there is a possibility that it is substantially different.

**Low certainty:** our confidence in the effect estimate is limited: the true effect may be substantially different from the estimate of the effect.

**Very low certainty:** we have very little confidence in the effect estimate: the true effect is likely to be substantially different from the estimate of effect.

**Explanations**

a. Three studies classified as some concerns of bias. Downgraded by one level.

b. The 95% confidence interval of two studies included a RR of 1. Downgraded by one level.

**Supplementary Figure S1.** Subgroup analysis for clinical remission (induction).

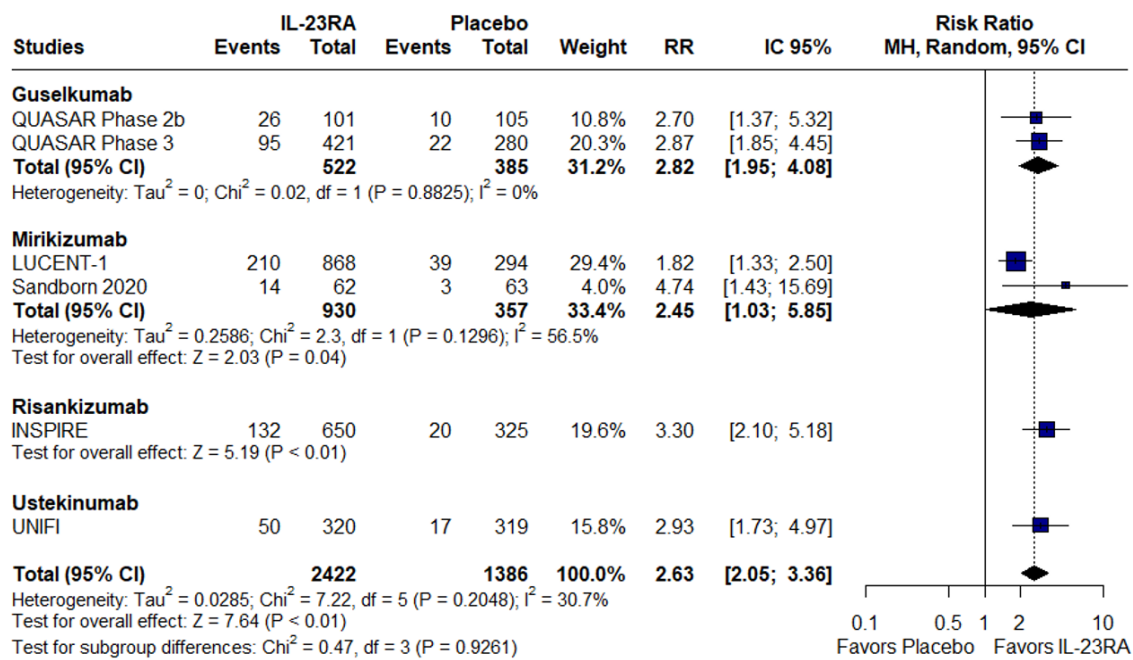

**Supplementary Figure S2.** Clinical remission in patients with previous failure with janus-kinase inhibitors or biologic therapy.

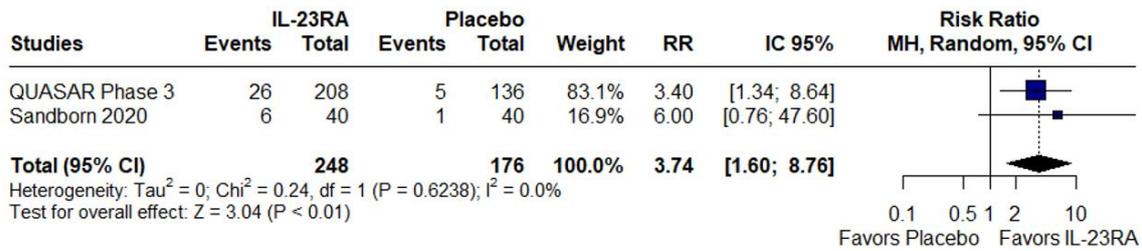

**Supplementary Figure S3.** Histologic, endoscopic, and mucosal response (induction).

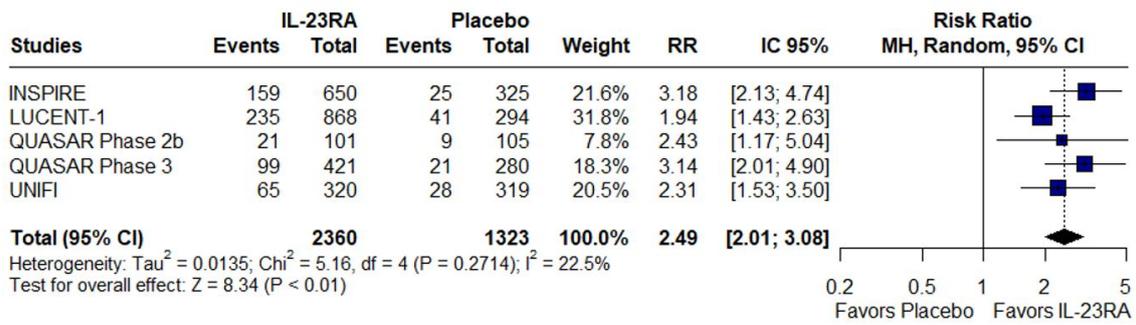

**Supplementary Figure S4.** Symptomatic remission (induction).

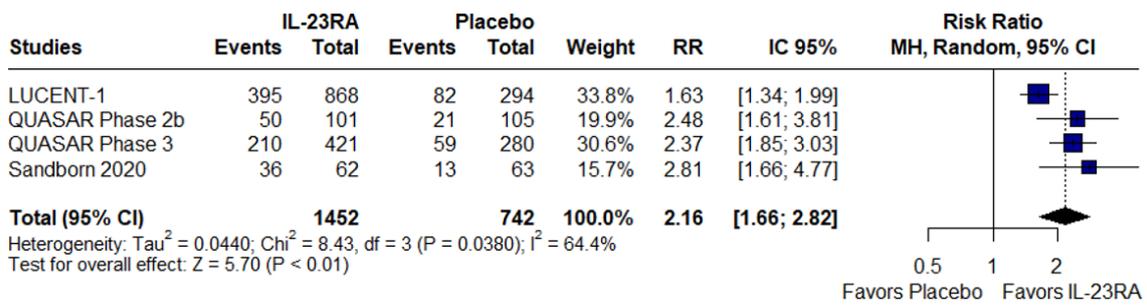

**Supplementary Figure S5.** Inflammatory Bowel Disease Questionnaire total score (IBDQ).

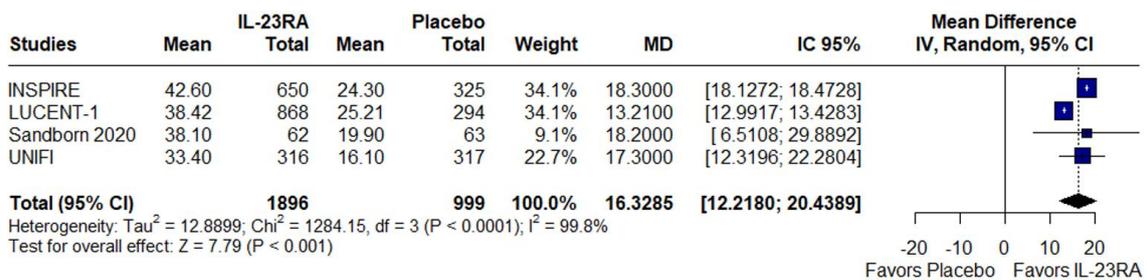

**Supplementary Figure S6.** Overall adverse events (induction).

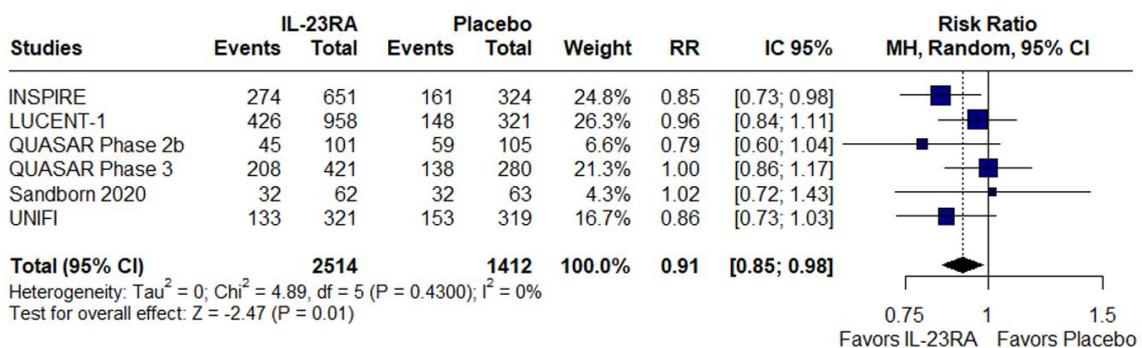

**Supplementary Figure S7.** Serious adverse events (induction).

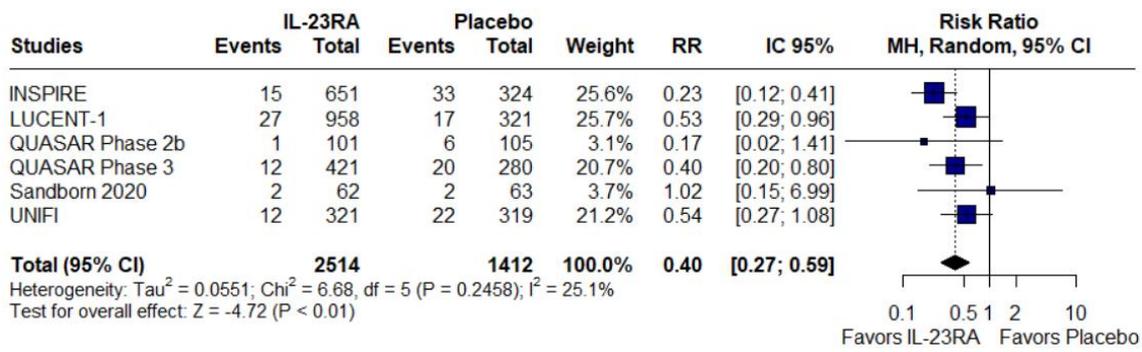

**Supplementary Figure S8.** Anemia (induction).

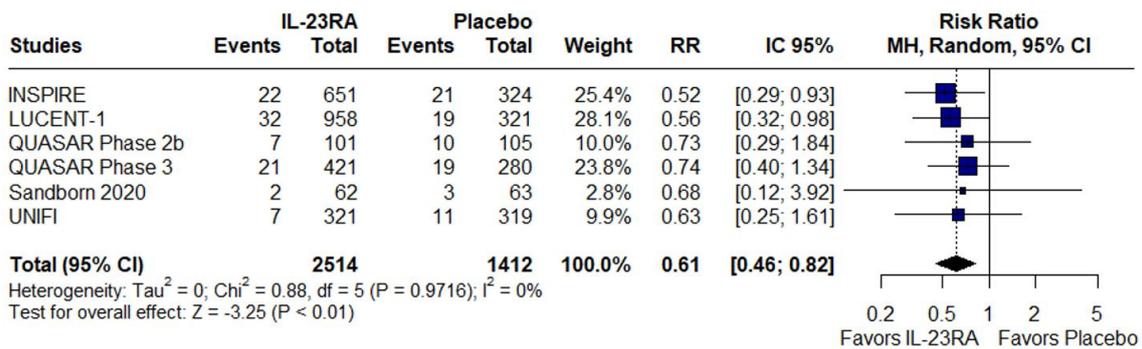

**Supplementary Figure S9.** Ulcerative colitis (induction).

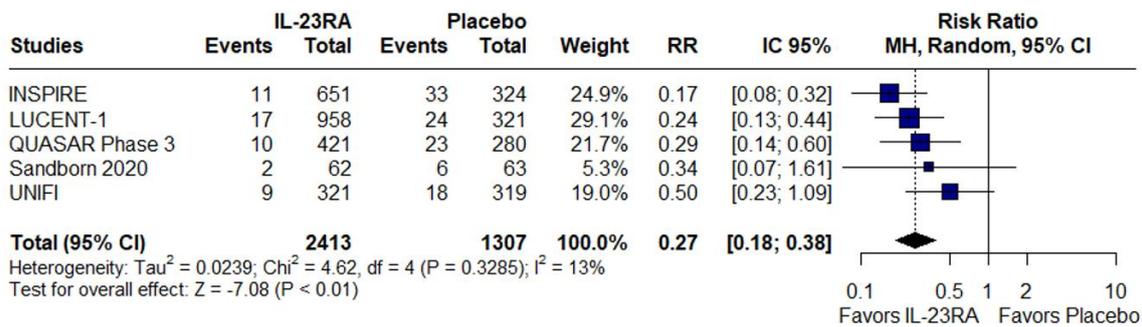

**Supplementary Figure S10.** Headache (induction).

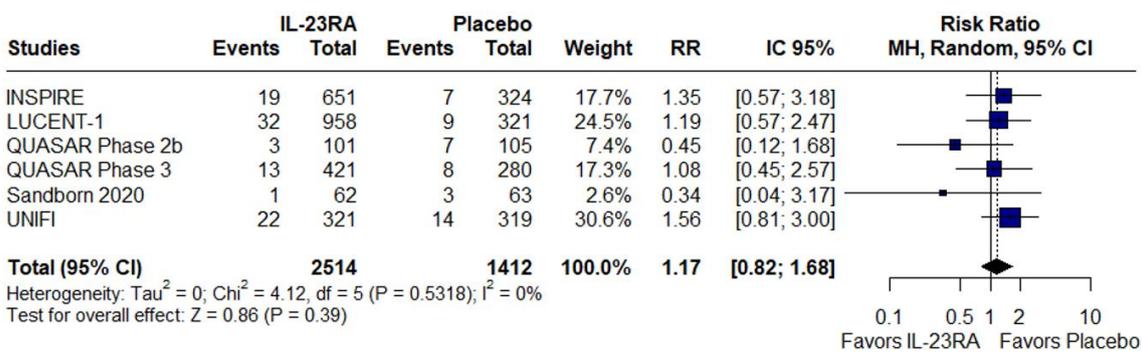

**Supplementary Figure S11.** Nasopharyngitis (induction).

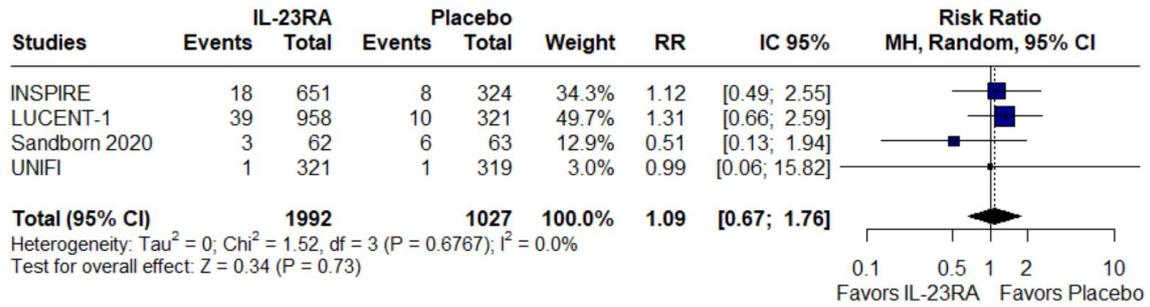

**Supplementary Figure S12.** Arthralgia (induction).

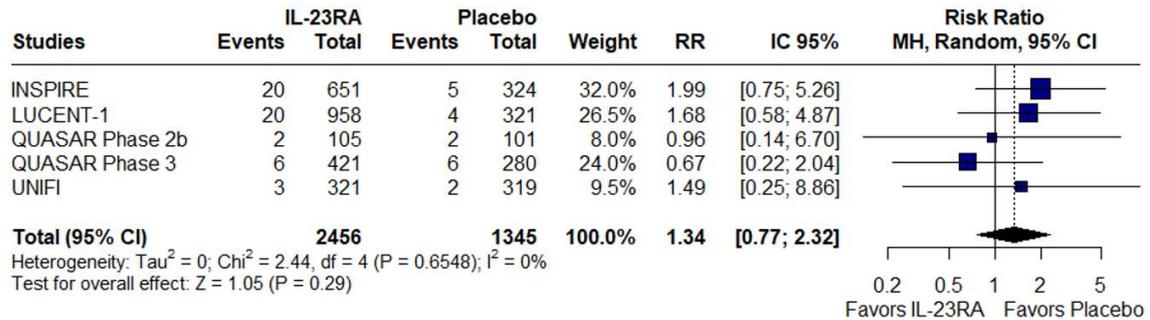

**Supplementary Figure S13.** Subgroup analysis for clinical remission (maintenance).

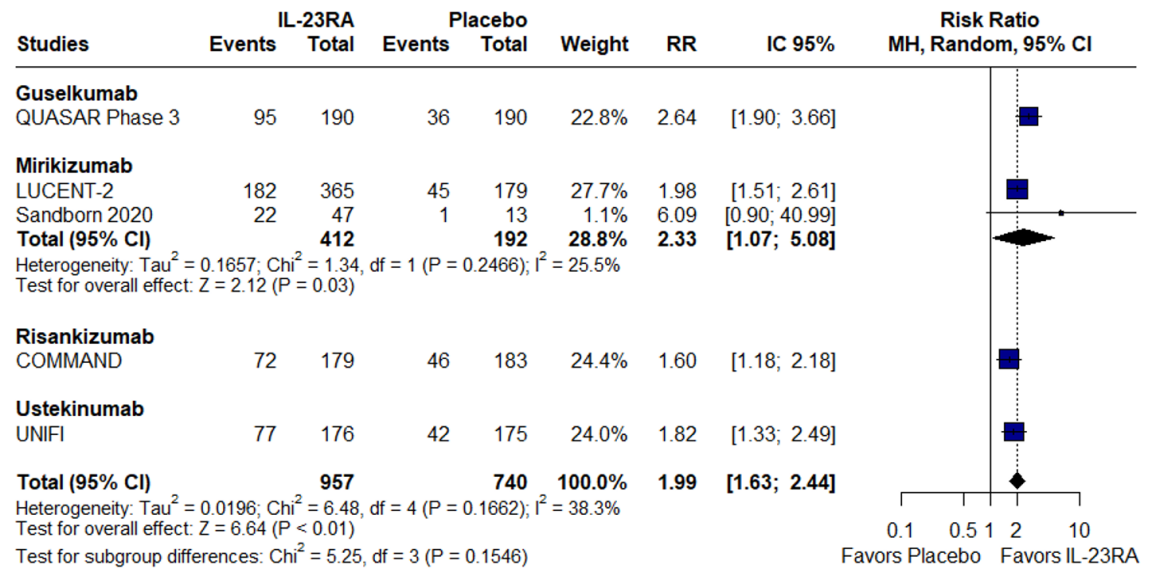

**Supplementary Figure S14.** Glucocorticoid-free remission (maintenance).

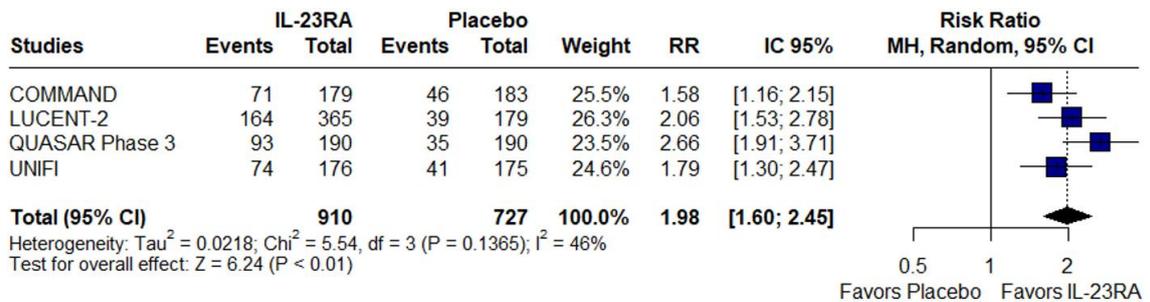

**Supplementary Figure S15.** Maintenance of clinical remission.

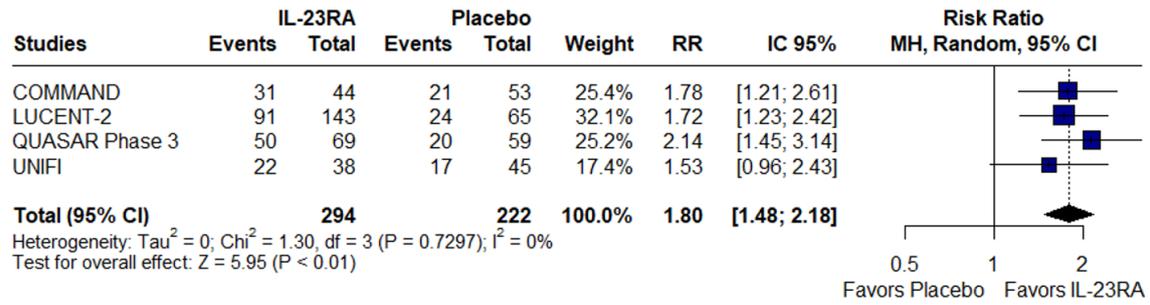

**Supplementary Figure S16.** Overall adverse events (maintenance).

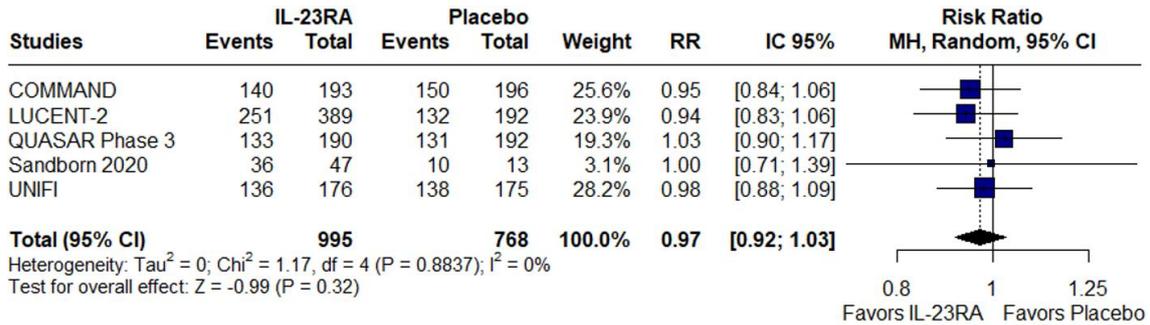

**Supplementary Figure S17.** Nasopharyngitis (maintenance).

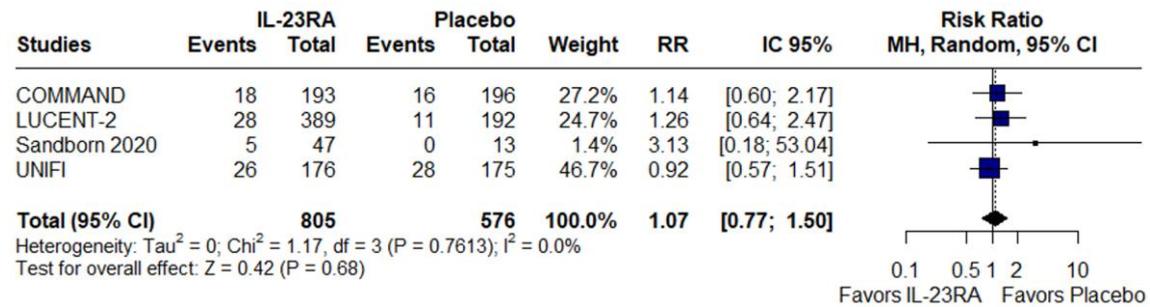

**Supplementary Figure S18.** Arthralgia (maintenance).

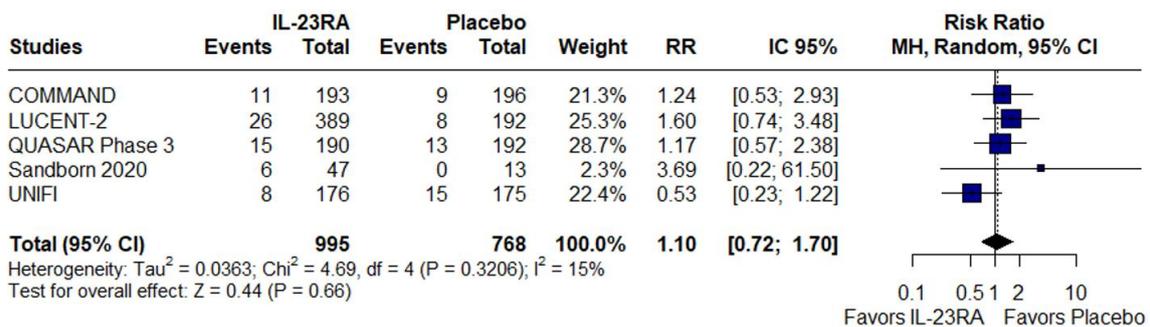

**Supplementary Figure S19.** Headache (maintenance).

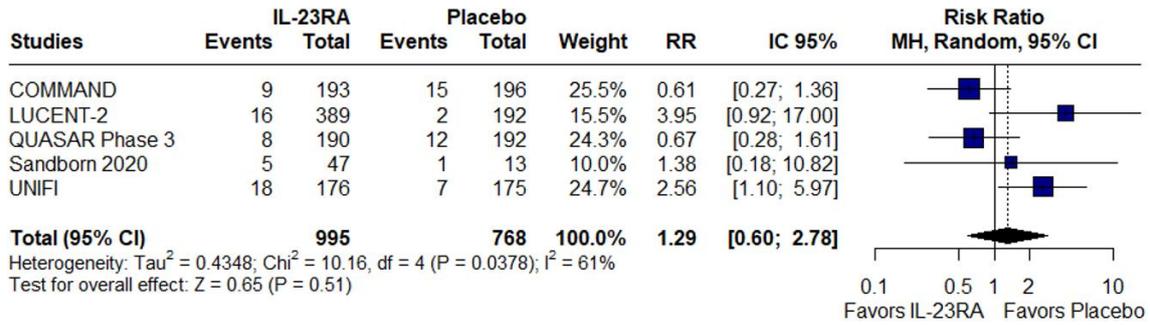

**Supplementary Figure S20.** Serious adverse events (maintenance).

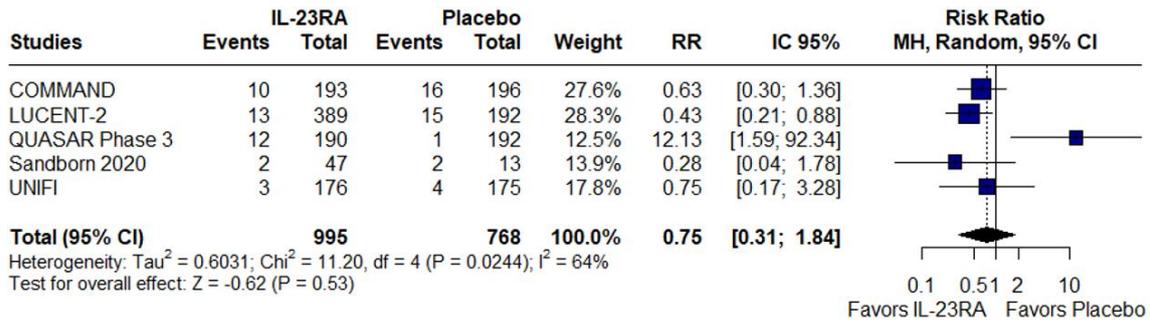

**Supplementary Figure S21.** Anemia (maintenance).

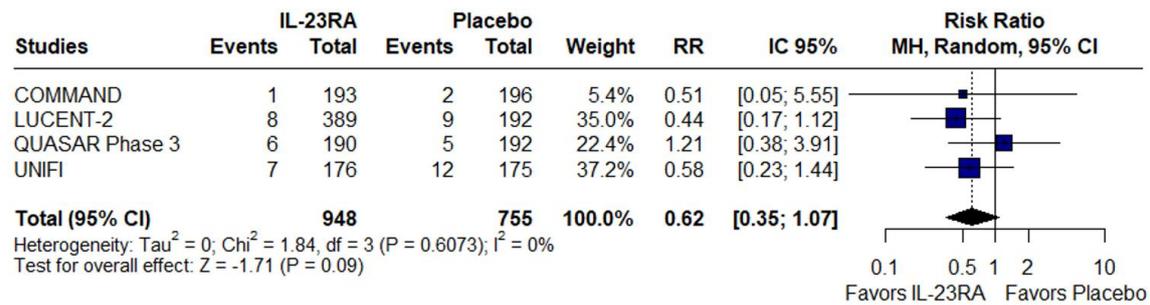

**Supplementary Figure S22.** Ulcerative colitis (maintenance).

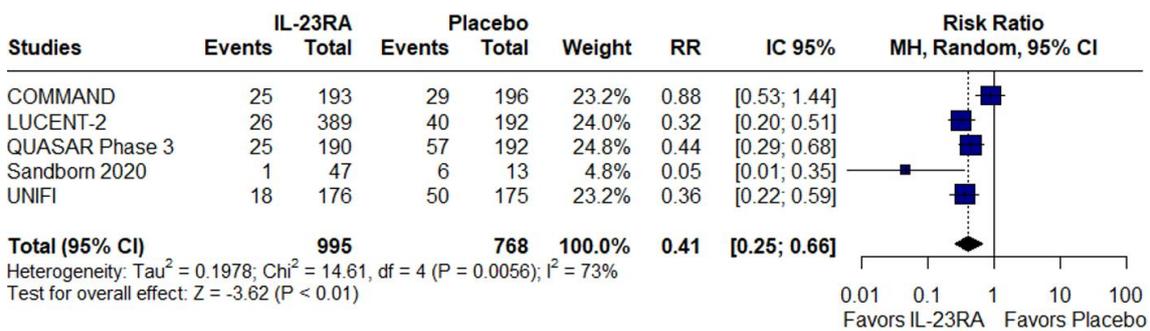

**Supplementary Figure S23.** Risk of bias assessment.

Risk of bias domains

| Study | D1 | D2 | D3 | D4 | D5 | Overall |
|---|---|---|---|---|---|---|
| COMMAND 2024 | + | − | + | + | + | − |
| INSPIRE | + | + | + | + | + | + |
| LUCENT-1 2024 | − | + | + | + | + | − |
| LUCENT-2 2024 | + | + | + | + | + | + |
| QUASAR Phase 2b 2023 | + | + | + | + | + | + |
| QUASAR Phase 3 Induction 2024 | + | + | + | + | + | + |
| QUASAR Phase 3 Maintenance 2024 | + | + | + | + | + | + |
| Sandborn 2020 | − | + | + | + | + | − |
| UNIFI 2019 | + | + | + | + | + | + |

Domains:
D1: Bias arising from the randomization process.
D2: Bias due to deviations from intended intervention.
D3: Bias due to missing outcome data.
D4: Bias in measurement of the outcome.
D5: Bias in selection of the reported result.

Judgement

− Some concerns

+ Low

**Supplementary Figure S24.** Funnel plot for clinical remission (induction).

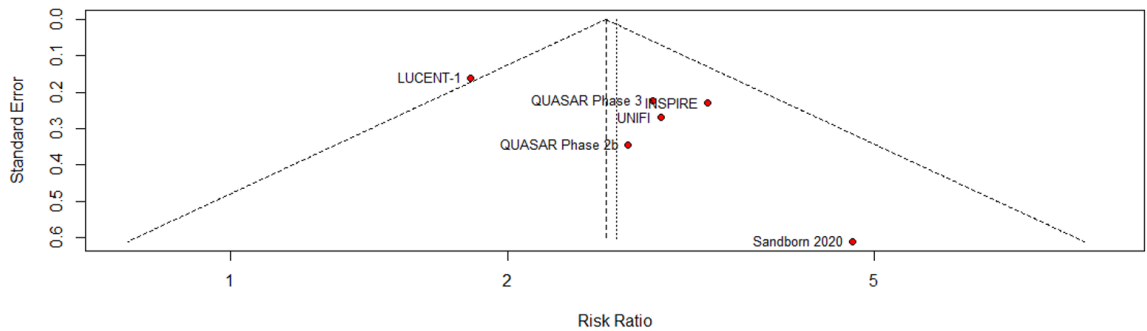

**Supplementary Figure S25.** Funnel plot for clinical response (induction).

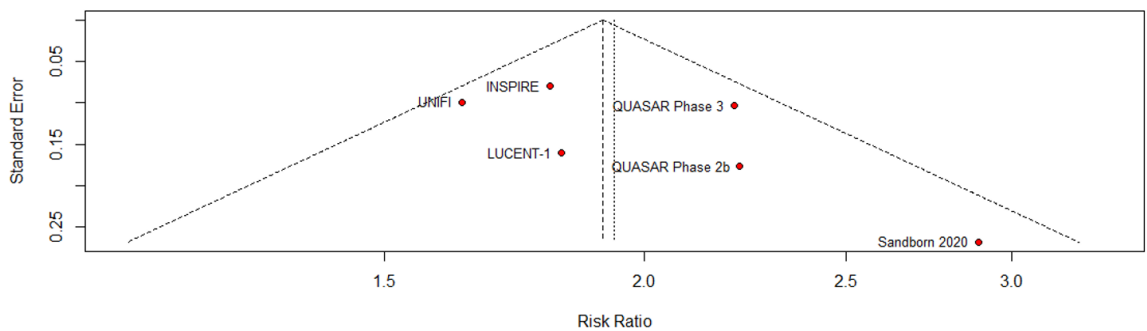

**Supplementary Figure S26.** Funnel plot for clinical remission (maintenance).

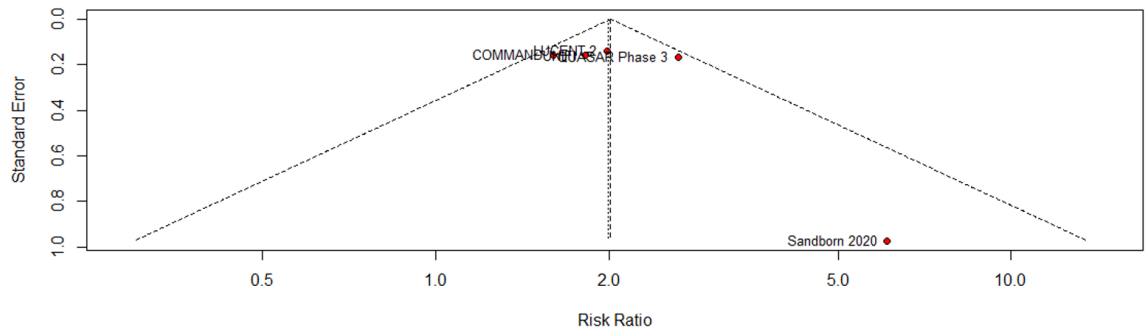

**Supplementary Figure S27.** Funnel plot for clinical response (maintenance).

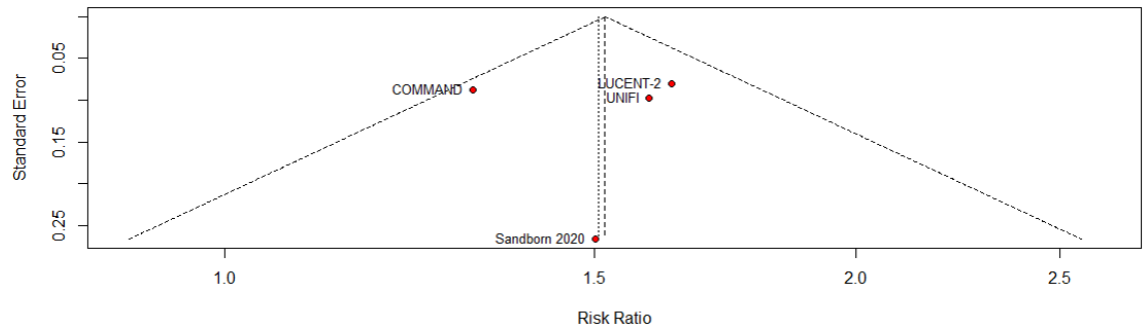

**Supplementary Figure S28.** Sensitivity analysis for clinical remission (induction).

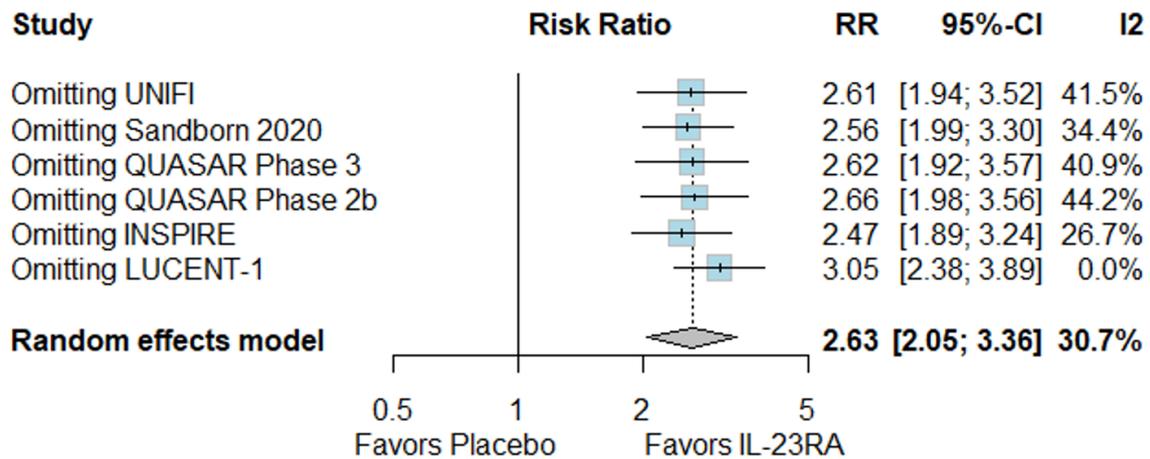

**Supplementary Figure S29.** Sensitivity analysis for clinical response (induction).

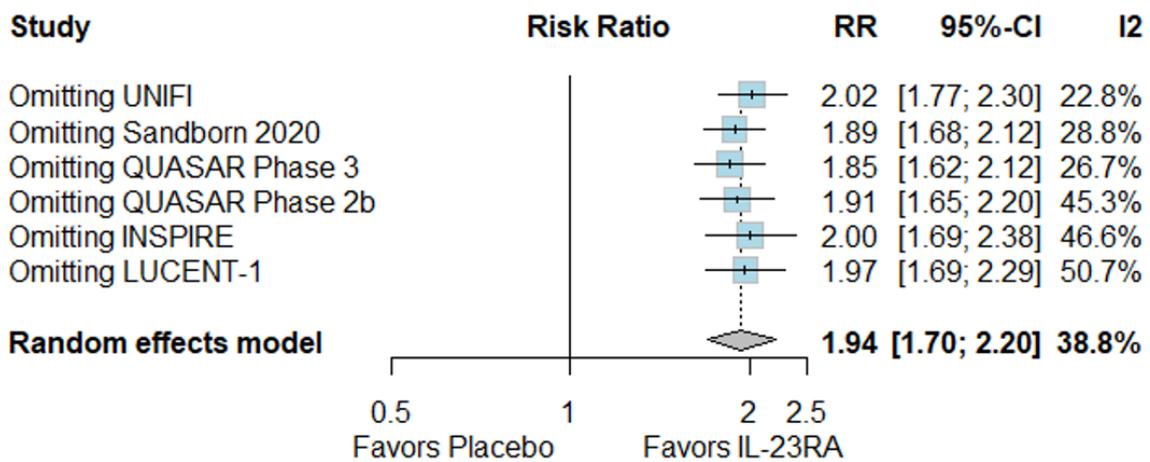

**Supplementary Figure S30.** Sensitivity analysis for clinical remission (maintenance).

| Study | Risk Ratio | RR | 95%-CI | I2 |
|---|---|---|---|---|
| Omitting UNIFI | | 2.07 | [1.57; 2.72] | 51.2% |
| Omitting Sandborn 2020 | | 1.97 | [1.61; 2.40] | 41.0% |
| Omitting QUASAR Phase 3 | | 1.83 | [1.54; 2.17] | 0.0% |
| Omitting COMMAND | | 2.13 | [1.72; 2.64] | 26.1% |
| Omitting LUCENT-2 | | 2.02 | [1.50; 2.72] | 53.9% |
| Random effects model | | 1.99 | [1.63; 2.44] | 38.3% |

1        5
Favors Placebo    Favors IL-23RA

**Supplementary Figure S31.** Sensitivity analysis for clinical response (maintenance).

| Study | Risk Ratio | RR | 95%-CI | I2 |
|---|---|---|---|---|
| Omitting UNIFI | | 1.47 | [1.25; 1.74] | 41.0% |
| Omitting Sandborn 2020 | | 1.51 | [1.31; 1.73] | 47.7% |
| Omitting COMMAND | | 1.61 | [1.43; 1.81] | 0.0% |
| Omitting LUCENT-2 | | 1.44 | [1.26; 1.65] | 10.4% |
| Random effects model | | 1.51 | [1.34; 1.69] | 21.5% |

1        2
Favors Placebo    Favors IL-23RA

**Supplementary Figure S32.** Bubble plot for meta-regression for outcomes of (A) clinical remission and (B) clinical response at induction phase according to the proportion of patients with prior failure to biologic therapy.

**A. Bubble plot for meta-regression for the outcome of clinical remission (induction)**

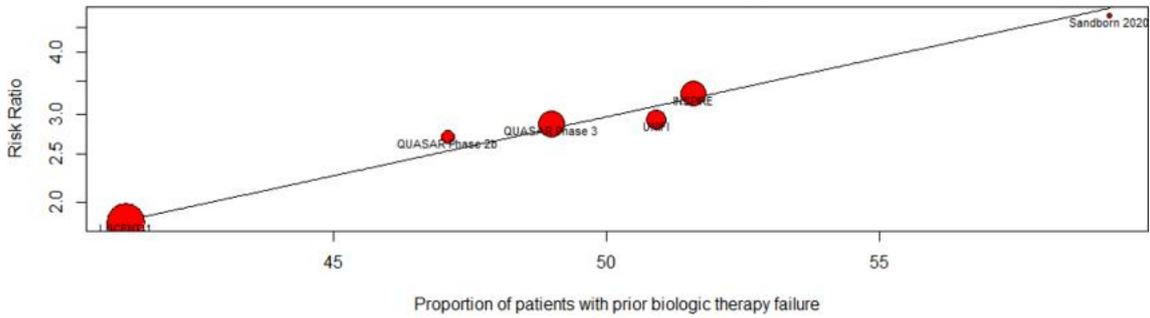

| | Effect Estimate | p-value | I² | Test for Residual Heterogeneity |
|---|---|---|---|---|
| Intercept | -1.63 | 0.09 | 0% | P = 0.99 |
| Biologic failure | 0.05 | 0.008 | | |

**B. Bubble plot for meta-regression for the outcome of clinical response (induction).**

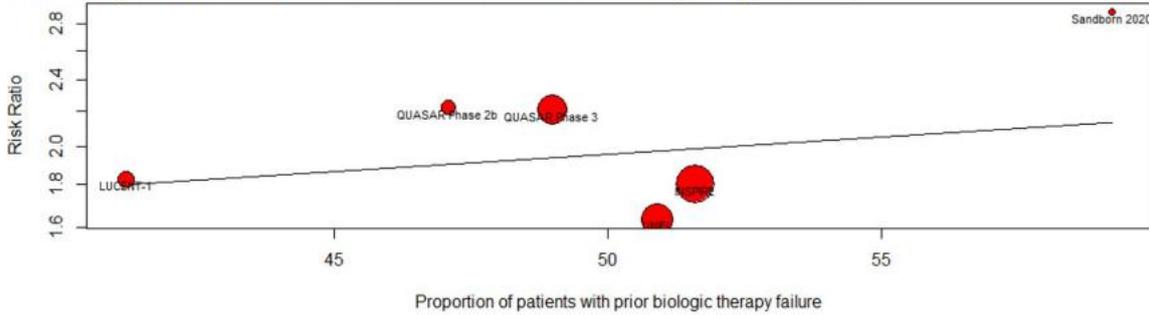

| | Effect Estimate | p-value | I² | Test for Residual Heterogeneity |
|---|---|---|---|---|
| Intercept | 0.19 | 0.82 | 51% | P = 0.09 |
| Biologic failure | 0.009 | 0.59 | | |

**Supplementary Figure S33.** Trial sequential analysis for clinical remission (induction).

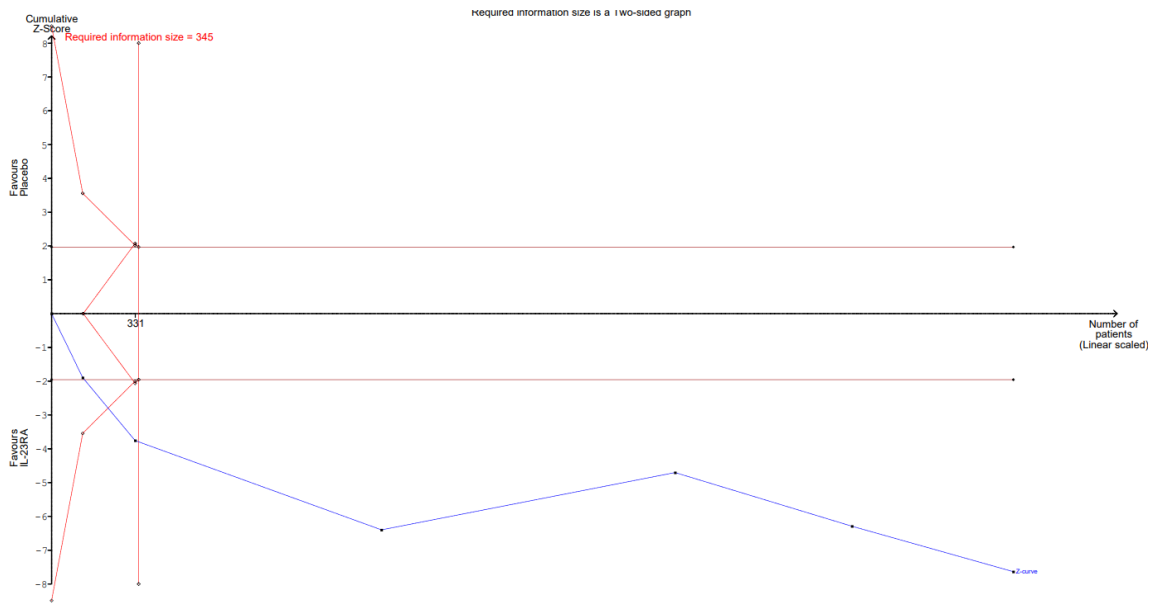

**Supplementary Figure S34.** Trial sequential analysis for clinical remission (maintenance).

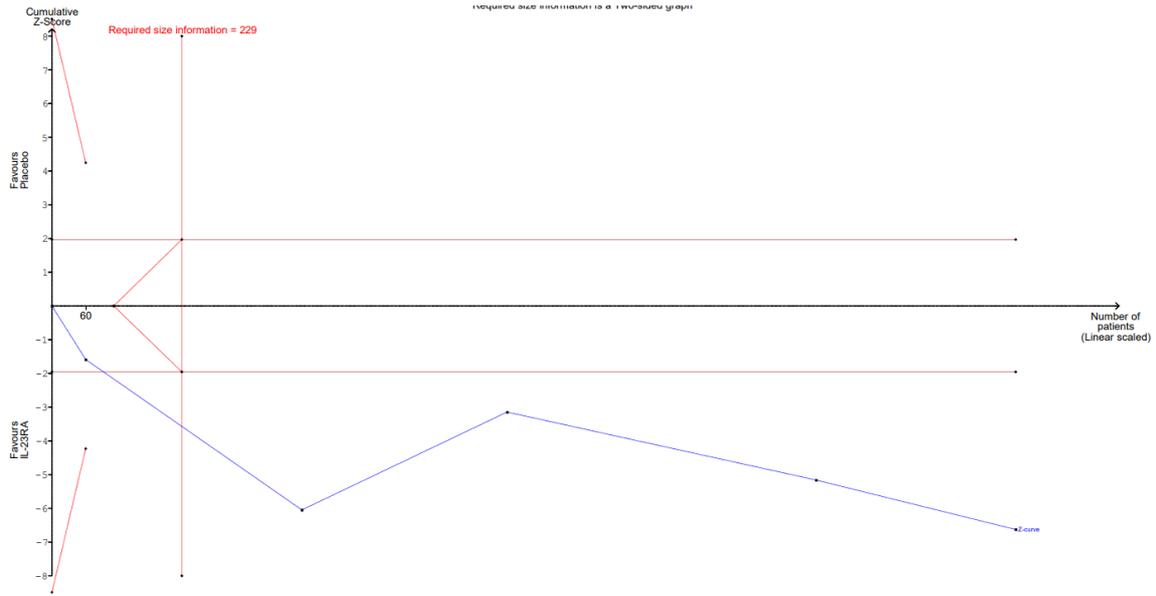

**Supplementary Figure S35.** Trial sequential analysis for clinical response (induction).

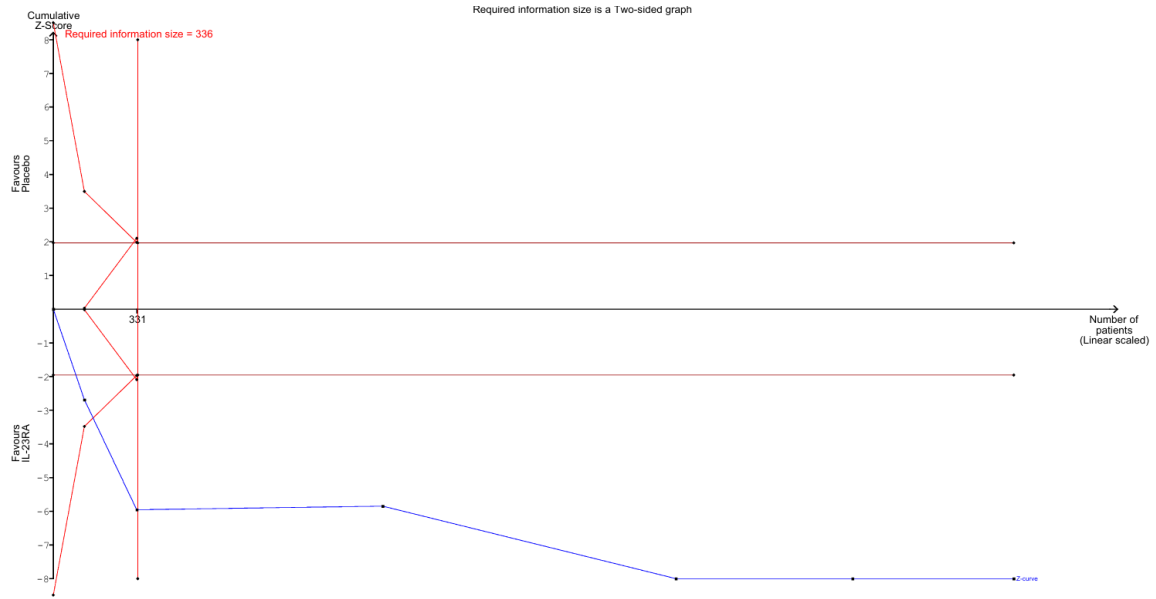

**Supplementary Figure S36.** Trial sequential analysis for clinical response (maintenance).

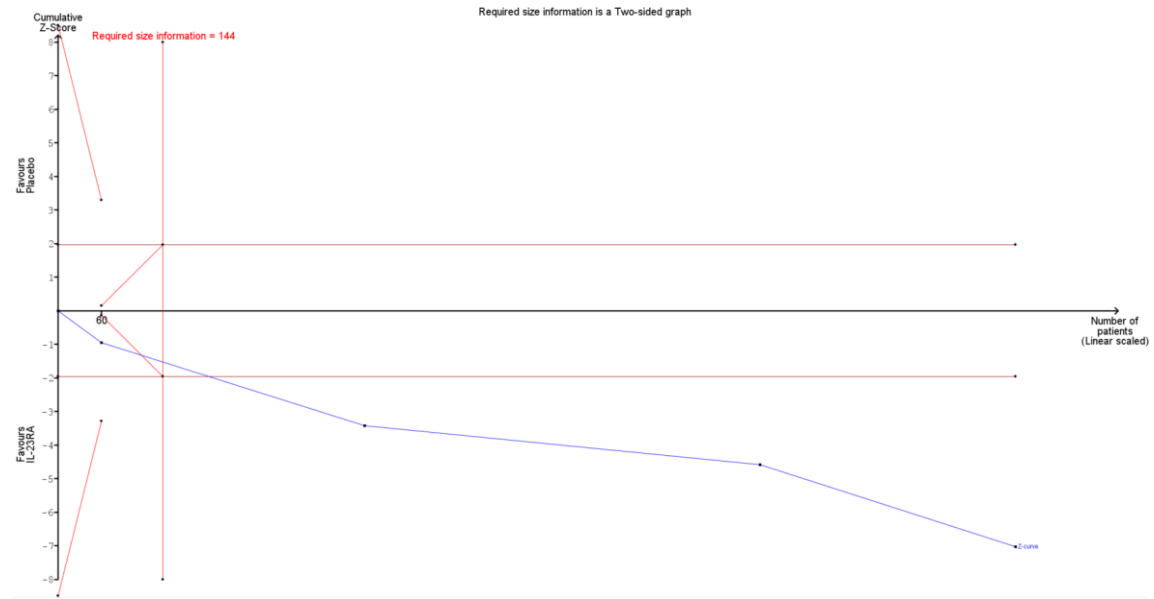